\documentclass[usenatbib]{mnras}
\usepackage[T1]{fontenc}
\DeclareRobustCommand{\VAN}[3]{#2}
\let\VANthebibliography\thebibliography
\def\thebibliography{\DeclareRobustCommand{\VAN}[3]{##3}\VANthebibliography}
\usepackage{amsmath}	
\usepackage{amssymb}	
\usepackage{comment}
\usepackage{array}
\usepackage{float}
\usepackage{bigints}
\usepackage{mathtools}
\usepackage{xfrac}
\usepackage[dvipsnames]{xcolor}
\usepackage{multicol} 
\usepackage{multirow}
\usepackage{subcaption} 
\usepackage{booktabs}
\usepackage{arydshln}
\usepackage{tabularx,colortbl}
\usepackage{makecell}
\usepackage{anyfontsize}
\usepackage[mathscr]{euscript}
\usepackage{dashrule}
\DeclareSymbolFont{rsfs}{U}{rsfs}{m}{n}
\DeclareSymbolFontAlphabet{\mathscrsfs}{rsfs}

\newcommand{\bc}{}

\newcommand{\avg}[1]{\langle{#1}\rangle}

\newcommand{\msunH}{${h^{-1}\rm M_\odot}~$}
\newcommand{\lsunhh}{{h^{-2}\rm L_\odot}}
\newcommand{\lsunhH}{{h^{-2}\rm L_\odot}~}
\newcommand{\mpch}{${h^{-1}\rm Mpc}$}
\newcommand{\mpcH}{${h^{-1}\rm Mpc}$~}
\newcommand{\chisq}{$\chi^2 \,$}
\newcommand{\redmapper}{{\sc redMaPPer}~}
\newcommand{\esD}{$\Delta\Sigma \mkern6mu $}
\newcommand{\vcutsymbol}{{\ooalign{\hfil$\vee$\hfil\cr\kern0.08em--\hfil\cr}}}

\defcitealias{Kumar2022}{KMR22}
\title[Subhalo masses of \redmapper satellites]{Environmental dependence on galaxy-halo connections for satellites using HSC weak lensing }
\author[ Amit Kumar et al.]{
Amit. Kumar,$^{1}$\thanks{E-mail: amitk@iucaa.in (IUCAA)}
Surhud More,$^{1, 2}$\thanks{E-mail: surhud@iucaa.in (IUCAA)}
\\
$^{1}$ Inter-University Centre for Astronomy and Astrophysics, Post bag 4, Ganeshkhind, Pune 411007, India\\
$^{2}$ Kavli Institute for the Physics and Mathematics of the Universe (WPI), 5-1-5 Kashiwanoha, 2778583, Japan
}
\date{Accepted XXX. Received YYY; in original form ZZZ}
\pubyear{2024}
\begin{document}
\label{Pub3_firstpage}
\pagerange{\pageref{Pub3_firstpage}--\pageref{Pub3_lastpage}}
\maketitle
\begin{abstract}
We present the luminosity-halo mass relations of satellite (sLHMRs) galaxies in the SDSS \redmapper cluster catalogue and the effects of the dense cluster environment on subhalo mass evolution. We use data from the Subaru Hyper Suprime-Cam survey Year-3 catalogue of galaxy shapes to measure the weak lensing signal around these satellites. This signal serves as a probe of the matter distribution around the satellites, thereby providing the masses of their associated subhalos. We bin our satellites based on physical observable quantities such as their luminosity or the host cluster's richness, combined with their cluster-centric radial separations. Our results indicate that although more luminous satellites tend to reside in more massive halos, the sLHMRs depend on the distance of the satellite from the cluster centre. Subhalos near the cluster centre (within $<0.3$\mpch) are stripped of mass. Consequently, the ratio of subhalo mass to luminosity decreases near the cluster centre. For low luminosity galaxies ($L < 10^{10} \lsunhh$), the lack of evidence of increasing subhalo masses with luminosity shows the impact of tidal stripping. We also present stellar-to-subhalo mass relations (sSHMRs) for our satellite sample evolving at different cluster-centric separations. Inferred sSHMRs in the outer radial bin appear to match that observed for the field galaxies. We show that the sSHMRs from the mock-\redmapper run on galaxy catalogues generated by the empirical UniverseMachine galaxy formation model are in good agreement with our observational results. Satellites, when binned based on the host cluster's richness, show very little dependence of the subhalo mass on the richness. 
\end{abstract}
\section{Introduction}
In the hierarchical model of structure formation, galaxies form within dark matter haloes. In this model, smaller haloes merge with larger ones and accrete matter from their surroundings \citep{Frenk2012}. Galaxies also co-evolve with their haloes. Post mergers, central galaxies become satellite galaxies within larger halos. These satellites sit in subhaloes of their own within the main bigger halo and correspond to the substructure within the halo. Massive haloes can host multiple galaxies, forming galaxy groups and clusters. The evolution of satellite galaxies in these clusters is influenced by a variety of processes, unlike field galaxies, which behave more like `island universes' \citep{Tormen_1998, Tormen2004, Lucia2004, Gao2004}. Physical processes like ram-pressure stripping can remove the hot gas from the galaxy, hence affecting its star formation rate \citep{Gunn1972}. Such physical processes can also alter the galaxy morphology due to frequent encounters with other satellites, a process known as galaxy harassment \citep{Moore1996, Moore1998}.

Satellite galaxies evolving in such dense environments experience dynamical friction, which causes their orbits to decay over time, leading them to sink toward the centre of the host's potential \citep{Binney2008}. Tidal effects felt by satellites are stronger near the cluster centre and die down with increasing separation from the cluster centre \citep{Springel2001, Lucia2004, Gao2004, Zhao2004, Xie2015}. One can study such effects by comparing the mass-to-light ratio of freshly accreted galaxies with earlier satellites \citep{Chang2013}. Cosmological simulations have been widely used to establish galaxy-halo connections for central as well as satellite galaxies \citep{Desmond2017, Bose2019, Girelli2020, Zhu2020, Bhattacharyya_2021, Yuan2022, Nadler2023, Sifon2024, Nusser2024, Puebla2024}. Collisionless dark matter simulations have revealed mass loss around satellites due to tidal stripping \citep{Gao2004, Bosch2005, Nagai2005, Giocoli2008, Xie2015, Rhee2017}. Hydrodynamical simulations can probe the effects of the dense environment on subhalo evolution \citep{Vogelsberger2014, Niemiec2019} in the presence of feedback processes, but are generally computationally expensive. 

Measuring the subhalo mass of the satellites is the biggest hurdle in quantifying the effects of dense environments on the evolution of galaxies. This is mainly because the dark matter does not interact electromagnetically. Since gravitational lensing is sensitive to the total mass of the lensing object, various studies in the literature have exploited this fact to compute the subhalo mass of the satellites. Multiple studies have quantified individual subhaloes of the lensing objects using flux perturbations in the strong lensing images \citep{Kneib1996, Koopmans2005, Vegetti2009, Vegetti2010, Kneib2011, Vegetti2012, He2022, Nightingale2022, He2023}. Although strong lensing can be used to estimate the subhalo masses, these instances are rare due to the low occurrence probability of such systems. Hence, quantifying subhalo masses for satellites evolving in galaxy groups and clusters remains challenging. 

The state-of-the-art technique to explore and analyse the subhalo masses of satellites is weak gravitational lensing \citep{Brainerd1996, Hoekstra2002, Mandelbaum2005, Mandelbaum2006, Yang2006, Poster_mira2011, Sifon2015,  Li2016, Sifon2018, Kumar2022, Wang2024}. In this technique, one measures the lensing shear imparted on the background source galaxies due to the satellite galaxies (which act as lenses). The lensing signal can then be used to infer the subhalo masses (for details, see Sec.\ref{Pub3_sec:weak_lensing_theory}). Independently, one can estimate the stellar masses of galaxies using spectral energy template fits to the photometry of galaxies to establish the satellite-subhalo relationship. 

In literature, multiple studies have established stellar-to-halo mass relations (hereafter SHMR) for central as well as satellite galaxies, exploiting the galaxy-galaxy lensing technique and satellite kinematics \citep{Hoekstra2005, heymans2006, Mandelbaum2006, Mandelbaum2016, More2011, van_uitert2011, van_uitert2016, Leauthaud2012, valander2014, zu2015, Coupon2015}. Recent developments in wide-field imaging surveys have made stacked galaxy-galaxy lensing measurements possible with unprecedented accuracy. \citet{Li2014} performed weak lensing measurements using imaging data from the Canada–France–Hawaii Telescope Stripe 82 (CS82) Survey \citep{Moraes2014} to constrain the subhalo masses of satellites from galaxy groups and clusters \citep{Li2016}. \citet{Sifon2015} used shape catalogue from the Kilo-Degree Survey \citep[KiDS,][]{de_Jong2013} to measure the weak lensing masses of subhaloes of satellite galaxies identified in Galaxy And Mass Assembly (GAMA) survey \citep{Driver2011}. They showed that satellites undergo significant mass loss compared to isolated field galaxies due to their evolution in a dense environment. On the contrary, on group scales, \citet{van_uitert2016} used satellites in GAMA groups with weak lensing from the KiDS to show that the mass-to-light ratio for satellite and field galaxies doesn't differ significantly. 

\citet{Niemiec2017} used combined shear catalogues from the Dark Energy Survey \citep[DES, ][]{Flaugher2015}, the CFHTLenS \citep{Heymans2012} and the CS82 to constrain stellar to halo mass relations for satellites evolving at different distances from the cluster centre. \citet{Sifon2018} used satellite galaxies from low redshift clusters defined in the Multi-Epoch Nearby Cluster Survey \citep[MENeaCS,][]{Sand2012} to establish galaxy-subhalo connections. \citet{Dvornik2020} measured the galaxy-galaxy lensing signal from KiDS in the background of GAMA galaxies to constrain SHMRs for central and satellite galaxies and to show differences in the stellar mass halo mass relation in the field and clusters.

\citet[][hereafter KMR22]{Kumar2022}  used the shape catalogue from the Hyper Suprime-Cam survey \citet[HSC][]{aihara2017} to measure the subhalo masses of satellites defined in \redmapper cluster catalogue \citep{Rykoff2014} to explore the effects of the dense cluster environment on subhalo mass evolution of satellites compared to their counterpart galaxies evolving in the field environment. They used the difference in the weak lensing signal of satellites and field galaxies to put an upper limit on the orphan fraction of galaxies in clusters \citep[see also][ for comparison with simulations]{Kumar2024}. Lastly, \citet{Wang2024} used imaging data from the Dark Energy Spectroscopic Instrument (DESI) Legacy Imaging Survey \citep[DECals][]{Dey2019} with \redmapper satellites to explore SHMRs for satellites with their halo-centric radius. 

Given that the stellar mass of satellite galaxies depends upon various assumptions of stellar population synthesis models, while the luminosity of the galaxy is a more direct observable. In this paper we study the relation between subhalo mass and the satellite galaxy luminosity. We improve upon \citetalias{Kumar2022} by using the stacked galaxy-galaxy lensing measurements from the HSC Year-3 shape catalogue, which corresponds to an increase in area by a factor of 3 compared to the Year 1 data used in \citetalias{Kumar2022}. We will use satellites from \redmapper and study the effect of the dense environment on the subhalo masses and their dependence on the luminosity of these galaxies. In addition, we will also study the effects of the environment on satellites in galaxy clusters of different richness. For the purpose of comparison with the literature we will also present the stellar mass halo mass relation of satellite galaxies.

\begin{table}
 \centering
 \begin{tabularx}{0.95\columnwidth} {    
  | >{\centering\arraybackslash}X
  | >{\centering\arraybackslash}X 
  | >{\centering\arraybackslash}X | }
  \hline
  \hline
  Luminosity  & $ R_{\rm sat_0}$ &   \multirow{2}{*}{\# of Satellites} \\ 
  {[log($h^{-2}L_{\odot}$)]} & [$h^{-1} \rm  Mpc$] & \\[0.5ex]
 \hline \\[-1.5ex]
 \multirow{2}{*}{(9.2 - 9.8]}  & 0.1 - 0.35 & 1224 \\ &0.35 - 0.6 & 1342 \\[1.5ex]
 \multirow{2}{*}{(9.8 - 10.1]}  & 0.1 - 0.35 & 1128 \\ &0.35 - 0.6 & 1154 \\[1.5ex]
 \multirow{2}{*}{(10.1 - 11.0]}  & 0.1 - 0.35 & 866 \\ &0.35 - 0.6 & 953 \\[1.ex]
 \hdashline \\[-1.5ex]
 \multirow{1}{*}{(9.2 - 9.8]}  & 0.1 - 0.6 & 2566 \\[1.5ex]
 \multirow{1}{*}{(9.8 - 10.1]}  & 0.1 - 0.6 & 2282 \\[1.5ex]
 \multirow{1}{*}{(10.1 - 11.0]}  & 0.1 - 0.6 & 1819 \\[0.5ex]
 
 \hline 
 \hline \\[-2.5ex]
 
  Richness ($\lambda$) & $ R_{\rm sat_0}/h^{-1}\rm Mpc$& \# of Satellites \\[0.5ex]
 \hline
 \\[-2.0ex]
 \multirow{2}{*}{(20 - 24]}  & 0.1 - 0.35 & 742 \\ &0.35 - 0.6 & 877 \\[1.5ex]
 \multirow{2}{*}{(24 - 28]}  & 0.1 - 0.35 & 764 \\ &0.35 - 0.6 & 790 \\[1.5ex]
 \multirow{2}{*}{(28 - 35]}  &  0.1 - 0.35 & 574 \\ &0.35 - 0.6 & 607 \\[0.5ex]
 \hline
\end{tabularx}
\caption{This table presents information on satellite galaxies defined in the \redmapper cluster catalogue within the redshift range of $0.1< Z\leq 0.33$, overlapping with HSC Y3 footprints. The satellites are member galaxies of clusters whose one of the putative central galaxies is defined with centring probability P$_{\rm cen}$ exceeding 0.95. We list counts of satellites in bins of projected radial distances based on their comoving separations $R_{\rm sat_0}$ from the cluster centre, combined with luminosity or richness-dependent selection cuts.}
\label{Pub3_table:satellies_data}
\end{table}

This paper is structured as follows. First, in Section~\ref{Pub3_sec:data}, we outline the lens and source catalogues used for the weak lensing measurements. Section~\ref{Pub3_sec:weak_lensing_theory} describes the theory of galaxy-galaxy lensing, along with signal measurement and null tests that we carry out for the weak lensing signal. In Section~\ref{Pub3_sec:modelling}, we describe the theoretical framework and the procedure employed to fit the measured weak lensing signal for satellites. Following this, we present our findings in Section~\ref{Pub3_sec:results_and_discussion} and discuss their implications. Finally, we summarize our results in Section~\ref{Pub3_sec:summary}.

In this paper, we use the terms "halo" and "subhalo" to refer specifically to the dark matter halos associated with central and satellite galaxies, respectively. Throughout the paper, "r" will denote the three-dimensional radial distance from the centre of a satellite galaxy and its projection on the sky as "R". Throughout this analysis, we use flat $\Lambda$CDM cosmology with parameters: $\Omega_m= 0.315$, $\Omega_bh^2=0.02205$, $\sigma_8=0.829$, $n_s=0.9603$ and $h=0.67$ \citep{Plank2014}. 

 \begin{figure*}
    \includegraphics[width=0.95\textwidth]{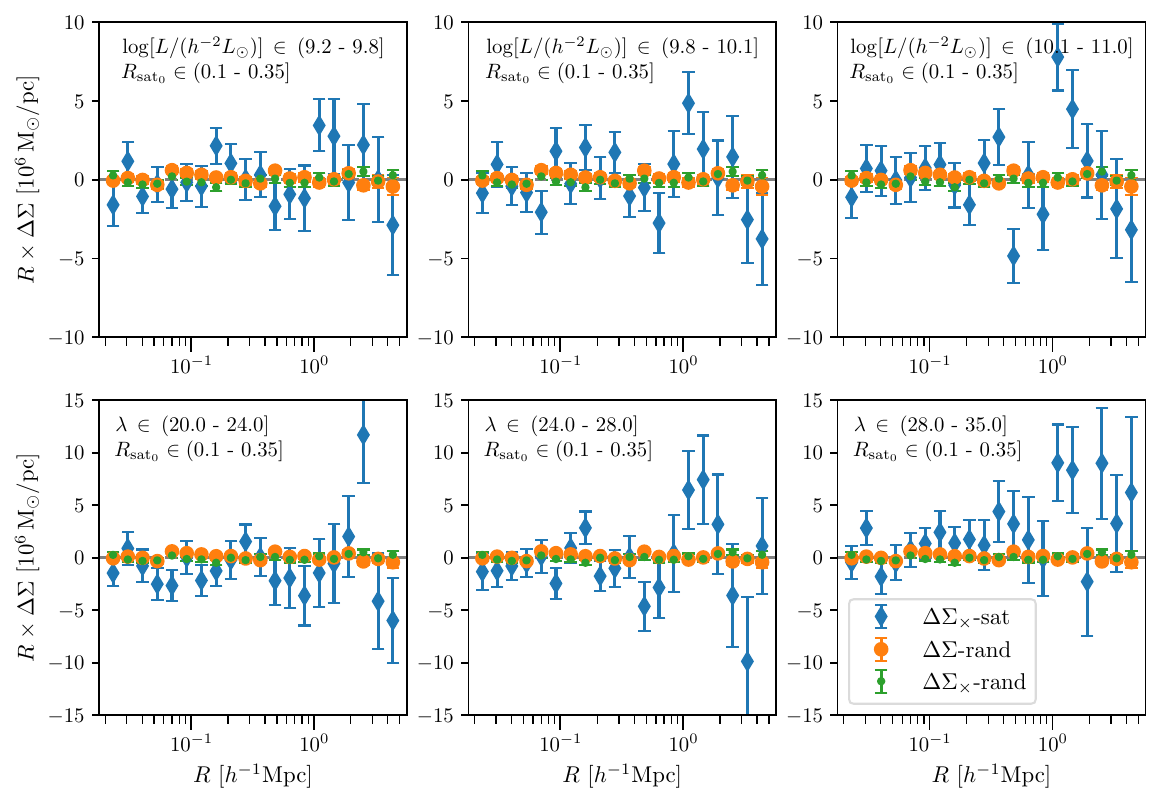}
    \caption{Systematics tests for different binning schemes: This plot is representative of the null tests on lensing signal measurements in different bins. The text in the bottom-left frame of each subfigure mentions the binning cuts on the sample. The diamond markers represent the cross component of the lensing signal, $\Delta\Sigma_{\rm X}$ around our lens galaxies from the inner radial bin, i.e. (0.1,0.35] in the luminosity (upper row) and richness (lower row) based binning schemes. The cross signal is consistent with zero for radial ranges, i.e. 0.02 to 0.5 \mpch, where the subhalo mass dominates. The orange and green colour data denote the tangential and cross components of the lensing signal for \redmapper randoms distributed in the same redshift range of our lens sample. }  
    \label{Pub3_fig:systematics_tests}
\end{figure*}

\begin{figure*}
	\includegraphics[width=0.98\textwidth]{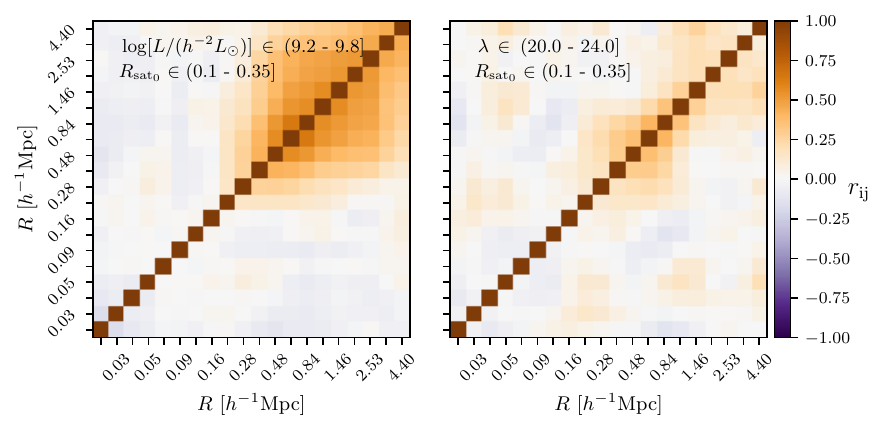}
	\caption{This figure shows the smoothed covariance matrices obtained using the Jackknife technique for one of the luminosity and richness bins combined with comoving projected radial separation selection cuts on satellites, i.e. $R_{\rm sat_0} \in $  (0.1,0.35] $h^{-1} \rm Mpc$ away from the cluster centre. The figure is representative of something similar we see for other bins too.}
    \label{Pub3_fig:covariance_matrices}
\end{figure*}

\section{Data}
\label{Pub3_sec:data}
This section outlines the data used in our analysis to measure the weak lensing signal around satellite galaxies. In particular, we use the satellite galaxies identified in the optically selected SDSS \redmapper cluster catalogue (described in Section~\ref{Pub3_sec:redmapper}) as lens galaxies. The high-quality images of galaxies identified from third-year data of the HSC-SSP survey serve as background source galaxies, as described in Section~\ref{Pub3_sec:hsc}

\subsection{SDSS \redmapper cluster catalogue} 
\label{Pub3_sec:redmapper}
The \textbf{red}-sequence \textbf{M}atched-filter \textbf{P}robabilistic
\textbf{P}ercolation (\redmapper) identifies galaxy clusters in photometric surveys as overdensities of red sequence galaxies \citep{Rykoff2014, Rykoff2016}. It uses photometric magnitudes and their errors to locate these over-densities of red sequence galaxies. In brief, it builds a model for the red sequence of galaxies based on their redshift, using a nominal training set of spectroscopic galaxies. The algorithm iteratively self-calibrates this model using multi-band photometric data and ultimately results in photometric redshift estimates with scatter ($\sigma_z/(1+z) \approx 0.01$ for z $\leq$ 0.7) for galaxy clusters \citep{Rykoff2016}. It assigns cluster membership probabilities to galaxies in the vicinity of the cluster to allow straightforward incorporation of uncertainty regarding whether a red galaxy is part of a specific cluster or not. The membership probabilities are further used to define the richness of the cluster. Comparisons with samples having spectroscopic redshifts indicate the memberships of satellites are accurate to 1\% and show low (<5\%) projection effects \citet{Rozo_2014}. Based on high-probability cluster members, the algorithm assigns a photometric redshift ($z_{\rm \lambda}$) to each cluster. In most of the clusters, a single bright galaxy is prominently positioned at the cluster's centre, while in some cases, it is not straightforward to identify a unique central galaxy unambiguously. Hence, \redmapper returns a list of five putative cluster centres from member galaxies with an associated probability $P_{\rm cen}$ of being the central galaxy, based on the overall radial, luminosity and redshift distribution of member galaxies.                

Over a redshift range of $0.1\leq z\leq 0.33$, where the incompleteness corrections due to the flux limit of SDSS are small, the \redmapper catalogue has more than $12\%$ clusters having richness $\lambda > 20$, where the probability of second galaxy being central is more than 30\%. Hence, to mitigate the issues related to miscentering, we exclusively select satellites only from clusters where the most probable central galaxy has a centring probability $P_{\rm cen} > 0.95$. In this analysis, we use galaxy clusters and randoms from the \redmapper catalogue v$6.3$ \citep{Rykoff2016} obtained from the SDSS DR8 \citep{Aihara2011} photometric galaxy catalogue. We restrict our lens sample of \redmapper satellites to the mentioned redshift range and to the region that overlaps with the footprint of the HSC year-3 shape catalogue, which we describe next.

\begin{table}
    \centering
    \begin{tabularx}{0.98\columnwidth}{
    >{\hsize=0.20\columnwidth\centering\arraybackslash}X
    >{\hsize=0.1\columnwidth\centering\arraybackslash}X
    >{\hsize=0.25\columnwidth\centering\arraybackslash}X
    >{\hsize=0.25\columnwidth\centering\arraybackslash}X
    }
    \hline\hline \\[-1.5ex]
    \multicolumn{2}{c}{}& \multicolumn{2}{c}{$\chi^2 \,{\rm for}\, \Delta\Sigma_{\times}$-{sat} } \\[0.75ex]
    \\[-1.5ex]    
    \multicolumn{1}{l}{}&\multicolumn{1}{l}{\textbf{$R_{\rm sat_0}\, \in$}}& \multicolumn{1}{c}{$ (0.1, 0.35]$} &\multicolumn{1}{c}{$(0.35, 0.6]$}  \\[0.75ex]
    \hline\\[-2.5ex]
    \multicolumn{2}{l}{log[Luminosity/($h^{-2}L_{\odot}$)]}&   \multicolumn{2}{c}{} \\    
     \multirow{1}{*}{(9.2-9.8]}  & & 10.0/12 (0.61) & 5.4/12 (0.94)\\
     \multirow{1}{*}{(9.8-10.1]} & & 9.9/12 (0.62) & 13.4/12 (0.34)\\
     \multirow{1}{*}{(10.1-11.0]} & & 16.5/12 (0.17) & 10.1/12 (0.6)\\ [0.8ex]
     
     \\[-0.8ex]
     \multicolumn{1}{c}{(9.2-9.8]}& &  \multicolumn{2}{c}{6.5/12 (.88)} \\
     \multicolumn{1}{c}{(9.8-10.1]}& &  \multicolumn{2}{c}{11.5/12 (0.49)} \\
     \multicolumn{1}{c}{(10.1-11.0]}& &  \multicolumn{2}{c}{14.2/12 (0.29)} \\
     \rule[0.1ex]{0.95\columnwidth}{0.1pt}\\[0.5ex]
     \multicolumn{2}{l}{Richness ($\lambda$)}& \multicolumn{2}{c}{} \\[1.ex]    
     
    \multirow{1}{*}{(20.0-24.0]} & & 16.4/12 (0.173) &15.2/12 (0.23)\\
    \multirow{1}{*}{(24.0-28.0]} & & 15.4/12 (0.22) & 6.2/12 (0.9)\\
    \multirow{1}{*}{(28.0-35.0]}& & 10.9/12 (0.53) & 13.9/12 (0.30)\\
    \hline 
    \hline \\[-1.5ex]    
    \multicolumn{1}{l}{}&\multicolumn{1}{l}{}& \multicolumn{1}{c}{$\Delta\Sigma $-{rand} } &\multicolumn{1}{c}{$\Delta\Sigma_{\times}$-{rand} }  \\[0.75ex]
    \multicolumn{2}{l}{Randoms}& 19.6/12(0.07) & 10.48/12(0.57)\\[1.ex]    
    \hline
    \end{tabularx}
    \caption{The table lists the systematics test statistics for the lensing signal measurements in different bins. The \chisq for the cross signal, along with the degrees of freedom for satellites in luminosity/richness bins combined with cluster-centric radial separation cuts, is provided for different combinations used in our analysis. The p-values to exceed \chisq corresponding to these results are listed in brackets alongside.}
    \label{Pub3_table:systematics}    
\end{table}

\subsection{ HSC-Y3 shape catalogue}
\label{Pub3_sec:hsc}
The HSC-Year 3 (hereafter Y3) galaxy shape catalog is based on the S19A internal data release from the HSC-SSP survey \citep{aihara2017, Aihara2018, aihara2019}, which includes data taken from March 2014 to April 2019 by the Hyper Suprime-Cam wide field camera on the 8.2 meter Subaru telescope \citep{miyazaki2012, Miyazaki2015, miyazaki2017, komiyama2017, furusawa2017, Kawanomoto2018}. This 870 Megapixels camera on the telescope has a 1.5 deg diameter field of view. The Y3 shape catalogue consists of 35.7 million galaxies spanning over an area of approximately $\sim$433 deg$^2$ in the northern sky and is divided into six disjoint fields (i.e. XMM, GAMA09H, WIDE12H, GAMA15H, VVDS, \& HECTOMAP). HSC is the deepest among the ongoing weak lensing surveys. The excellent i-band median seeing of $\sim$ 0.59, high galaxy number density, wide field of view, and a large sky area make HSC a unique survey to carry out weak lensing studies. 

Various selection cuts were used to limit the galaxies from the S19A data for conducting weak lensing analyses as described in \cite{Li2022}. These include restricting galaxies to the Full Depth Full Color region of the survey, i.e. region which reaches approximately full survey depth in all five (g, r, i, z, y) broadband filters. This results in uniform quality for the shape and photometric redshift measurements of all galaxies in the catalogue. The catalogue is further restricted to galaxies having an $i-$band c-model magnitude i$_{\rm cmodel}$  brighter than 24.5, i-band SNR$\geq10$, > 5$\sigma$ detection in at least two bands other than i, in order to have galaxies whose shapes can be reliably measured, and such that their redshifts can be calibrated reliably with the help of existing spectroscopic or multi-band photometric redshift data.

The galaxy shapes in the observed galaxy sample are determined with the help of the re-Gaussianization (\texttt{reGauss}) shape measurement method on the $i-$band coadded images \citep{Hirata2003}. This method measures two components of galaxy ellipticity as
\begin{equation}
    (e_{\rm 1},e_{\rm 2}) = \frac{1-(b/a)^2}{1+(b/a)^2}(\cos{2\psi},\sin{2\psi}).
    \label{Pub3_eq:e1e2}
\end{equation}
Here, $b/a$ signifies the observed minor-to-major axis ratio, while $\psi$ denotes the position angle formed by the major axis of the source galaxy relative to the equatorial coordinate system. 
Each source galaxy is assigned a shape weight $w_{\rm s}$, based on its shape measurement error $\sigma_{\rm e}$, and the intrinsic root mean square ellipticity per component $e_{\rm rms}$, given as
\begin{equation}
w_{\rm s}=(\sigma_{\rm e}^2+e_{\rm rms}^2)^{-1}.
\label{Pub3_eq:shape_weight}
\end{equation}
The biases in the shape measurements were calibrated using image simulations with a methodology similar to \citet{Mandelbaum2018b}. The simulations use Hubble Space Telescope images from the COSMOS region degraded to resemble the image quality expected from ground-based imaging in the HSC survey and to match the distribution of properties of the HSC galaxies \citep{Li2022}. Subsequently, the multiplicative and additive biases in shape measurements for shear estimation were determined for each object by carrying out the shape measurements on sheared copies of simulated galaxies. \citealt{Li2022} estimated that the uncertainties in the overall shear calibration are of order $10^{-2}$. The null tests conducted on the shear catalogue illustrate that the systematic uncertainties in it are sufficiently small to enable cosmological analysis using galaxy clustering galaxy-galaxy lensing and cosmic shear \citep{Dalal2023, Li2023, Miyatake2023, More2023, Sugiyama_2023}. Despite implementing relatively stringent cuts, the HSC third-year shear catalogue results in a high number density, i.e. 22.9 $\rm arcmin^{-2}$ (raw) and 19.9 $\rm arcmin^{-2}$ (effective) of galaxies with $i_{\rm cmodel} < 24.5$, with a median redshift of 0.8 \citep{Li2022}. This makes the Y3 shape catalogue ideal for studying the galaxy-galaxy lensing of redMapper satellites, which we restrict to redshifts less than $0.33$.

The primary contribution to systematics in weak lensing signal measurements arises from uncertainties associated with photometric redshift (photo-z) estimates. Errors in photometric redshift estimation may result in foreground lens galaxies being erroneously classified as part of the source galaxy population, which can dilute the weak lensing signal. Given the high number density of source galaxies and the survey depths, it is practically impossible to obtain spectroscopic redshifts for all source galaxies, hence the use of photometric redshifts is necessary. HSC Y3 survey provides photo-z estimates for galaxies based on three different codes: two based on deep neural networks (dNNz (Nishizawa et al. in prep), DEMPz \citep{Hseish2014, Tanaka2018}, and one spectral energy distribution template fitting, MIZUKI \citep{Tanaka2015}. We use the redshift PDFs from the photometry of galaxies inferred from the dNNz method to define our sample of source galaxies as well as to calculate the lensing signal for our satellites, as we describe in the following section. 

\section{Galaxy-galaxy lensing: Weak, Theory and measurements}
\subsection{Weak lensing: Theory}
\label{Pub3_sec:weak_lensing_theory}
The coherent distortions imprinted on the images of background source galaxies due to the presence of intervening matter are known as weak gravitational lensing. These distortions are a result of convergence and shear, which cause a uniform magnification of the original image and deformations to it, respectively. In practice, the intrinsic shape of the galaxy is never known. Hence, it is impossible to find the magnification and distortions to it just from imaging data. However, it is possible to extract the lensing shear at different projected separations from the centre of the lensing object by averaging the shapes of many source galaxies lying behind it. Hence, the weak lensing shear is computed in a statistical manner. In the weak lensing regime, the expectation value of the ellipticity can be related to the lensing shear (\citealt{Schramm_Kayser_1995}, \citealt{Seitz_Schneider_1997}). One can use the measured ellipticities (see Eq. \ref{Pub3_eq:e1e2}) of source galaxies to estimate their tangential ellipticity, $e_{\rm t}= -e_{\rm 1} \cos(2\phi) - e_{\rm 2} \sin(2\phi)$, with respect to the line joining the source galaxy and the lens. This tangential ellipticity is related to the tangential lensing shear $\gamma_t$.

The strength of the lensing shear depends on the distance of the source and lens from the observer and is sensitive to the projected matter density along the line of sight. The average of observed quantity $\gamma_t$ is proportional to the excess surface mass density (ESD, i.e. $\Delta\Sigma$ ), also known as the lensing signal as
\begin{equation}
\avg{\gamma_{\rm t}(R)}=\frac{\Sigma(<R) -\Sigma(R)}{\Sigma_{\rm crit}(z_{\rm l},z_{\rm s})}=\frac{\Delta\Sigma(R)}{\Sigma_{\rm crit}(z_{\rm l},z_{\rm s})}\,.
\label{Pub3_eq:esd}
\end{equation}
Here, $\Sigma$(<$R$) refers to the average surface mass density within a projected distance $R$, and $\Sigma(R)$ denotes the azimuthally averaged surface mass density within a thin annulus at that distance $R$. Whereas,
$\Sigma_{\rm crit}(z_{\rm l},z_{\rm s})$ factor is given by
\begin{equation}
\Sigma_{\rm crit}(z_{\rm l},z_{\rm s})= \frac{c^2}{4\pi G}\frac{D_{\rm A}(z_{\rm s})}{D_{\rm A}(z_{\rm l}) D_{\rm A}(z_{\rm l},z_{\rm s})(1+z_{\rm l}^2)}\,,
\label{Pub3_eq:sigma_crit}
\end{equation}
and accounts for the distance of the source and lens from the observer. In this notation, $D_{\rm A}$ denotes the angular diameter distances, and $z_l$, $z_s$ represent redshifts of lens and source, respectively. The inclusion of an additional factor $(1+z_{\rm l})^2$ in the denominator is attributed to our adoption of comoving coordinates. \citep{Bartelmann_and_Schneider_2001}.

The HSC shape catalogue provides the information of the ellipticity ($e_{\rm 1},e_{\rm 2}$) and the shape weight $w_{\rm s}$ for each observed galaxy. The catalogue also provides information on the multiplicative bias $\hat{m}$ and additive biases ($c_{\rm 1},c_{\rm 2}$) for the galaxies based on simulation calibrations \citep{Mandelbaum2018b, Li2022}. We write the weighted signal for many-many lens-source galaxy pairs using Eq.~\ref{Pub3_eq:esd} as
\begin{equation}
\begin{split}
    \Delta\Sigma(R)&=\frac{1}{(1+\hat{m})}\left(\frac{\sum_{\rm ls}w_{\rm ls} e_{\rm t,ls}\left\langle\Sigma_{\rm crit}^{-1}\right\rangle^{-1}}{2\mathcal{R}\sum_{\rm ls}w_{\rm ls} }\right.\\&\qquad\qquad\qquad\qquad\qquad\left.- \frac{\sum_{\rm ls}w_{\rm ls} c_{\rm t,ls}\left\langle\Sigma_{\rm crit}^{-1}\right\rangle^{-1}}{\sum_{\rm ls}w_{\rm ls} } \right),
\end{split}    
\label{Pub3_eq:stacked_signal}
\end{equation}
at a radial separations $R$, from the lens center. Where, $\langle\Sigma_{{\rm crit}}^{-1}\rangle$ is defined using probability distribution for the source galaxy redshift $P(z_{\rm s})$ as
\begin{equation} 
    \langle\Sigma_{\rm crit}^{-1}\rangle=\frac{4\pi G(1+z_{\rm l})^2}{c^2}\int_{z_{\rm l}}^{\infty}\frac{D_{\rm A}(z_{\rm l}) D_{\rm A}(z_{\rm l},z_{\rm s})}{D_{\rm A}(z_{\rm s})} P(z_{\rm s})dz_{\rm s}.
\end{equation}

We use only source galaxies for which posterior redshift distribution satisfies, 
\begin{equation}
\int_{z_{l,\rm max}+z_{\rm diff}}^{\infty}P(z_s) dz_{s} \geq 0.99,
\label{Pub3_eq:ellipticities}
\end{equation}
where we set  $z_{l,\rm max}$ to the maximum redshift in our lens galaxy sample, i.e. \redmapper satellites. This condition implies there is a more than 99\% probability for the source galaxy in consideration of being at a higher redshift than all the galaxies in the lens sample. The extra separation in the redshift of $z_{\rm diff}$=0.1 helps make our source selection further robust. Also, the term $w_{\rm ls}=w_{\rm s}\langle\Sigma_{\rm crit}^{-1}\rangle^2$ dilutes the lensing signal contributed from the closely separated lens-source galaxy pairs. The factor $\hat{m} \,(=\Sigma_{\rm ls}w_{\rm ls}m_s/\Sigma_{\rm ls}w_{\rm ls}$)  in the Eq.~\ref{Pub3_eq:stacked_signal}, represents multiplicative bias. Lastly, $\mathcal{R}$ in this equation is known as shear responsivity and quantifies how a small applied shear will affect the ellipticity measurements \citep{Bernstein_Jarvis_2002}. It is computed using RMS intrinsic shape distortions $e_{\rm rms}$, combined with $w_{\rm ls}$ as, 
\begin{equation}
    \mathcal{R}=1- \frac{\sum_{\rm ls}w_{\rm ls }e_{\rm RMS}^2}{\sum _{\rm ls} w_{\rm ls}}.
\end{equation}

\subsection{Weak lensing: Measurements \& Systematics}
\label{Pub3_sec:observations}
In our analysis, satellite galaxies defined in the optically selected \redmapper cluster catalogue serve as lens samples. We would like to investigate the effects of the galaxy cluster environment on the mass-luminosity relationship for satellites. Hence, we adopt a luminosity-based binning scheme for them,  combined with their cluster-centric radial separations. For this purpose, we use the rest frame luminosity of satellites, obtained by applying k-correction \citep{Blanton2007} to their observed R-band model magnitude, `MODEL\_MAG\_R' provided in the catalogue. In particular, in this binning scheme we measure the weak lensing signal for satellites having comoving projected radial cluster centric separation $R_{\rm sat_0}$ $\in$ (0.1-0.35], (0.35-0.6] $h^{-1}$Mpc, combined with luminosity bins of (9.2,9.8],(9.8,10.0] and (10.0,11.0] log($h^{-2}L_{\odot}$). 

In the \redmapper cluster catalogue, richness is a proxy of the population of member satellites. To study the effect of cluster richness on the subhalo mass of the satellites orbiting at different distances, we divide satellites in bins of host cluster's richness, i.e. (20,24],(24,28],(28,35] along with their projected radial separation (0.1-0.35], (0.35-0.6] $h^{-1}$Mpc from the cluster centre. 

In table:\ref{Pub3_table:satellies_data}, we list the counts of satellites, meeting selection cuts from these luminosity and richness-based binning schemes. We calculate the weak lensing signal for satellites in 20 logarithmic bins evenly spaced between 0.02 - 5 \mpcH from the centre of the satellite. 

We perform two null tests on the measured lensing signal, i.e. measurement of the cross component of lensing signal $\Delta\Sigma_{\times}$, and signal around random lenses. The $\Delta\Sigma_{\times}$ is obtained by averaging the cross component of lensing shear, which in turn is a $45^{\circ}$ rotated ellipticity vector with respect to the major/minor axis of the source galaxy, taking reference to the line joining the source galaxy to the centre of lensing object. We performed the second null test, i.e. \esD measurements around the random points, using \redmapper randoms as lens galaxies. In the absence of systematics, both these measurements should be consistent with zero within the statistical uncertainty.

The upper and lower rows in Fig. \ref{Pub3_fig:systematics_tests} are representative of null tests on the measured signal for luminosity and richness-based binning schemes, respectively. The selection cuts based on these schemes are indicated in the left-bottom text frame of each sub-figure. In this figure, the orange and green circles denote the lensing signal and its cross-component for random points, whereas the diamond markers in blue represent measurements of the cross-component of the lensing signal for satellites from the inner radial bin. The \chisq for random points following the same distribution as that of galaxies is $\sim$ 24 for 20 degrees of freedom (dof), corresponding to a p-value of 0.24. The cross signal for random points has \chisq $\sim$ 21 for 20 dof with a p-value of 0.39. Given the uncertainties, the \esD and $\Delta\Sigma_{\times}$ for random points is consistent with zero. We find the cross signal for satellites has high \chisq for some of the bins. But, in the range radial range $R \in [0.02,0.5]$ \mpcH, which is the region that dictates the subhalo mass, this is also within the acceptable range. Here, we have shown systematics plots only for the inner radial bin, i.e. $R_{\rm sat_0}$ $\in$ (0.1-0.35), which is also representative of the outer radial bin. Nevertheless, in Table  \ref{Pub3_table:systematics}, we list systematics for all the bins used in our analysis. The systematics at large values of R in some of the bins are somewhat high, but these scales don't affect our inference of subhalo masses of satellites. In the table, we list the p-values to exceed reduced \chisq for the region, which dictates the subhalo masses, i.e. $0.02-0.5$ \mpcH. 

\subsection{Covariance }
\label{Pub3_sec:covariance}
The non-zero variance of the ensemble average of the intrinsic ellipticities of the background source galaxies due to their finite number results in the presence of shape noise in the weak lensing signal. The covariance resulting from the shape noise can be estimated by randomly rotating all the source galaxies. The shape noise term scales inversely with the $\sqrt N$, where N is the number of lens-source galaxy pairs. It dominates the error budget on the measurements of lensing signal for small values of projected distance $R$ due to the smaller number of lens-source pairs available at these distances. At large values of $R$, lens-source pairs increase due to the increased area of the radial bin. Hence, the covariance due to the shape noise term decreases. The error budget on these scales has a significant contribution from the large-scale structure, as it can coherently shift the measurements of the lensing signal up or down. The \esD measurements in different radial bins can be correlated as the same source galaxy may contribute to the lensing signal around multiple lens galaxies at different distances.

We address the above-mentioned sources of noise collectively by using the Jackknife technique \citep{Miller1974}. We divide the entire HSC Y3 survey area into 256 rectangular approximately equal area regions, using the HSC random catalog distributed in the same area as our lens galaxies. We then measure the lensing signal around our satellites, excluding one region at a time from the entire dataset. We compute the covariance matrix $\mathbf{C}$ as,
\begin{equation}
\begin{split}
    C_{ij} &=\frac{N_{\rm jack}-1}{N_{\rm jack}}\sum_{m=1}^{N_{\rm jack}}\Big[ \Big( \Delta\Sigma(R_{i,m}) -\overline{\Delta\Sigma}(R_{i}) \Big)  \\&\qquad\qquad\qquad\qquad\qquad  \Big( \Delta\Sigma(R_{j,m}) -\overline{\Delta\Sigma}(R_{j})\Big) \Big]. 
\end{split}
\label{Pub3_eq:correlation_matrix}
\end{equation}
Here, the indices i and j vary from 1 to 20, corresponding to logarithmic radial bins of projected distance $R$ used in lensing signal computations. Whereas $\overline{\Delta\Sigma}(R_i)$ is obtained by averaging the lensing signal in all the jackknife regions for $i^{th}$ radial bin. The cross-correlation matrix  $r_{ij}$, defined as  
\begin{equation}
r_{ij}=\frac{C_{ij}}{\sqrt{C_{ii}C_{jj}}},
\label{Pub3_eq:correlation_coeff}
\end{equation}
with the help of $C_{ij}$ component of covariance matrix $\mathbf{C}$. The left and right panels of Fig. \ref{Pub3_fig:covariance_matrices} are representative of smoothed correlation matrices obtained using Jackknife techniques for one of the luminosity and richness-based selection bins used in our analysis. To smoothen out noise in the covariance matrix, we smooth the cross-correlation matrix using a box-car filter of size 3 following the procedure adopted in \cite{mandelbaum2013}. The selection values used in both the sub-figures are indicated in respective upper-left text boxes. As expected, we observe less correlation in off-diagonal terms of the matrix at smaller values of projected distance $R$, and correlation increases at larger scales.  

\section{Model \& Fits}
\label{Pub3_sec:modelling}
\begin{table}
\renewcommand{\arraystretch}{1.5}
 \centering
 \begin{tabularx}{0.9\columnwidth} {    
  | >{\centering\arraybackslash}X 
   |>{\centering\arraybackslash}X | }
  \hline
   \multicolumn{2}{c}{Model Parameter Priors}\\[0.5ex]
  \hline
  Parameters&Flat Priors Range\\
  \hline
  $\log[M_{\rm clu}/h^{-1}M_{\odot}]$& [ 10 - 16 ] \\
  $\log[M_{\rm sub}/h^{-1}M_{\odot}]$& [ 9 - 16 ]\\
  $\log[M_{\rm bary}/h^{-1}M_{\odot}]$&[ 8 - 14 ]\\
  $\sigma_{\rm mc}$& ( 0 - 1 )\\[0.5ex]
 \hline
\hline
\end{tabularx}
\caption{The table lists the flat prior ranges used for sampling the posterior distribution of the free model parameters in our MCMC analysis. The parameters  $M_{\rm clu}$, $M_{\rm sub}$ represent the average sub(halo) masses of the main cluster and satellite galaxies, respectively, while $\rm M_{\rm bary}$, denotes the average baryonic mass of the satellite galaxies. Lastly, $\rm \sigma_{\rm mc}$ quantifies the width of the 2-D Rayleigh probability distribution used for the mis-centering analysis.}
\label{Pub3_table:priors}
\end{table}

\begin{figure*}
  \includegraphics[width=0.99\textwidth]{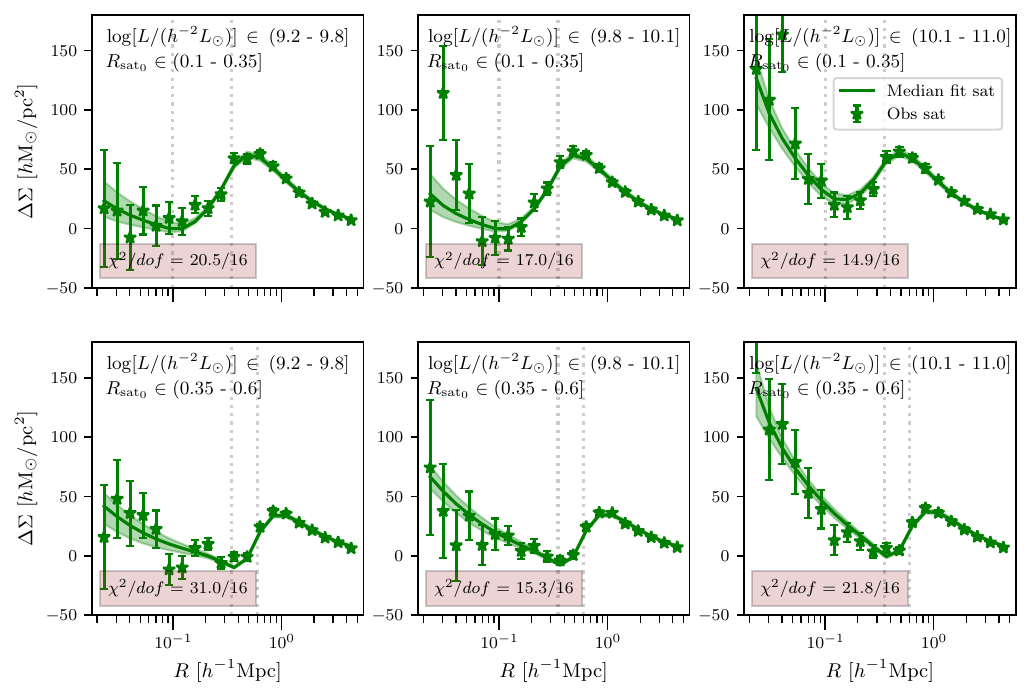}
  \caption{The markers in the figure show the galaxy-galaxy lensing signal for satellites as a function of satellite-centred on sky projected distance $R$. The upper and lower rows in the figure correspond to the satellite sample selected based on their luminosity but have different separations from the cluster centre (0.1,0.35],(0.35,6] \mpcH respectively. The text box in the top middle of each subfigure reflects different selections used on satellites in that particular analysis. The solid line in each subfigure represents the median model fitted to the lensing signal using parameters obtained from MCMC sampling. Whereas the goodness of the fit indicated by the $\chi^2/{\rm dof}$ is inscribed in the text box located at the lower-left position in each of the subfigures.} 
\label{Pub3_fig:signal_luminosity_bins}
\end{figure*}

\begin{figure*}
  \includegraphics[width=0.99\textwidth]{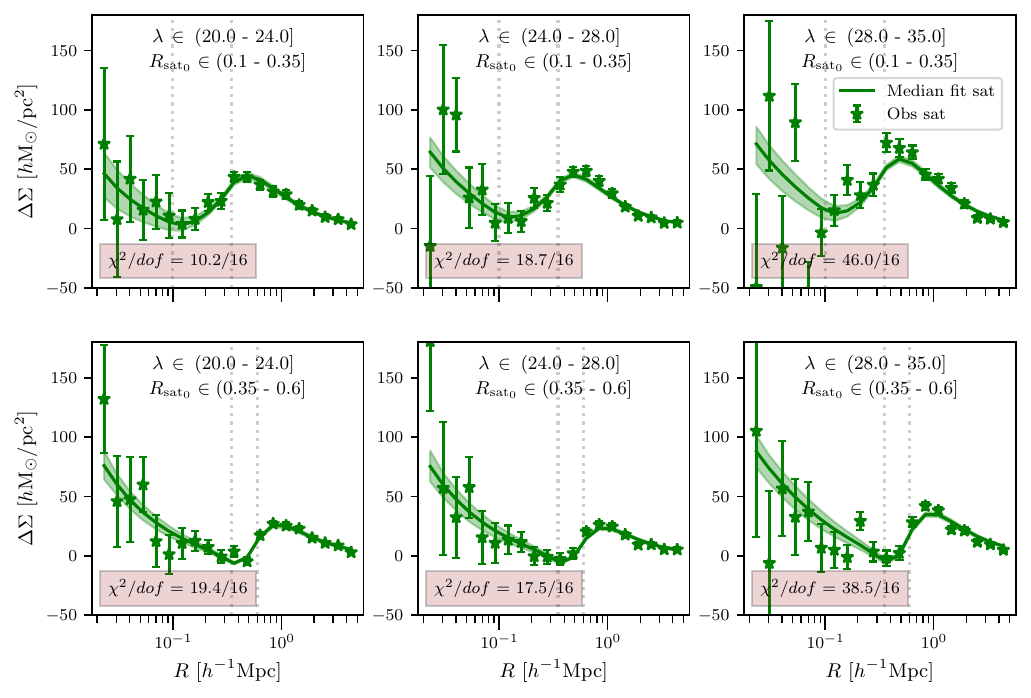}
  \caption{Similar to Fig. \ref{Pub3_fig:signal_luminosity_bins}, but satellites binned based on cluster richness rather than their luminosity.} 
  \label{Pub3_fig:signal_richness_bins}
\end{figure*}

The model used in this analysis is described extensively in Sec. 4 of \citetalias{Kumar2022}. We describe it here, in short, for the sake of completeness. We use this model to fit the measured weak lensing signal for satellites in luminosity or richness bins listed in Table \ref{Pub3_table:satellies_data}.

\subsection{The density profile}
\label{Pub3_sec:density_profile}
We use the Navarro, Frenk \& White \citep{Navarro1997} density profile to describe the dark matter distribution of the (sub)halo of satellites and their host BCG and integrate it along the line of sight to get the projected  surface density profile given by,
\begin{equation}   
\Sigma^{\rm DM}(R) = \int \rho_{\rm NFW}\left(\ \sqrt{R^2+z^2} \right) dz\,.
\label{Pub3_eq:sigma_R}
\end{equation}
Subsequently, the average surface mass density within a circle of a given radius $R$ is calculated as 
\begin{equation}
\Sigma^{\rm DM}(<R)=  \frac{2}{R^2} \int_{0}^{R} \Sigma^{\rm DM}(R') R' dR'.
\label{Pub3_eq:sigma_inside}  
\end{equation}
The above Eqs.\ref{Pub3_eq:sigma_R} \& \ref{Pub3_eq:sigma_inside} allow us to calculate the excess surface mass density using Eq.\ref{Pub3_eq:esd}. We use the analytical expression in Eq. 14 of \cite{Wright2000} to evaluate the lensing $\Delta\Sigma(R)$ at a projected distance, $R$, from the galaxy centre. 

\subsection{Model signal for satellite galaxies}
\label{Pub3_sec:subhalo_contribution}

The dark matter surrounding satellites within galaxy clusters can be expressed as the sum of two components: the dark matter associated with the halo of the BCG, and the subhalo mass of the satellite itself. The \esD for a satellite galaxy can be expressed as the sum of its baryonic and dark mass,
\begin{equation}
\Delta \Sigma(R) =  \Delta\Sigma^{\rm bary}_{\rm sat}(R)+\Delta \Sigma^{\rm DM}_{\rm sat} + \Delta \Sigma^{\rm DM}_{\rm clu}(R). 
\label{Pub3_eq:deltasigmatotal}
\end{equation}
The first term in this equation, $\Delta\Sigma^{\rm bary}_{\rm sat}(R)$ accounts for the baryonic mass of the satellite and drops as $R^{-2}$, i.e. $\Delta\Sigma^{\rm bary}_{\rm sat}(R)={M_{\rm bary}}/{4\pi R^2}$. Assuming the dark matter distribution of the subhalo of the satellite follows the NFW  density profile, we estimate the second term in Eq.\ref{Pub3_eq:deltasigmatotal}, $\Delta \Sigma^{\rm DM}_{\rm sat}$ by using Eq. 14 of \cite{Wright2000}, as discussed earlier in Sec.\ref{Pub3_sec:density_profile}.
In order to account for the matter contribution from the host cluster around the satellite, $\Delta \Sigma^{\rm DM}_{\rm clu}(R)$ correct determination of the cluster centre is crucial. Hence, we use satellites from only those clusters where the BCG is defined with a centring probability > 0.95. Nevertheless, we model  $\Delta \Sigma^{\rm DM}_{\rm clu}(R)$ with an off-centred NFW density profile for the BCG, to account for the miscentering of the cluster centre. We assume the true cluster centre follows a Rayleigh distribution  \citep{Johnston2007} around the \redmapper assigned cluster centre. In the parametric form of this probability distribution, \begin{equation}
    P(R_{\rm off})= \frac{R_{\rm off}}{2\pi \sigma_{\rm mc}^2}\exp{\left (- \frac{R_{\rm off}^2}{2\sigma_{\rm mc}^2}\right) }\,,
\end{equation}
the scale parameter $\sigma_{\rm mc}$ characterizes the width of the distribution, and we leave it as a free parameter in our model. Hence for a assigned central-satellite separation $R_{\rm sat_0}$, and free parameter $\sigma_{\rm mc}$, we can rewrite the $\Delta\Sigma^{\rm DM}_{\rm clu}$ as
\begin{equation}\begin{split}
\Delta\Sigma^{\rm DM}_{\rm clu}&(R|R_{\rm sat_0},\sigma_{\rm mc})=\\
&\quad \int_{0}^{2\pi}\int_{0}^{\infty} \Delta\Sigma^{\rm DM}_{\rm clu}(R|R^{\rm true}_{\rm sat})\; P(R_{\rm off}|{\sigma_{\rm mc}})\; dR_{\rm off}d{\theta}\,, 
\label{Pub3_eq:miscentering}
\end{split}\end{equation}
to account for the off-centring of the cluster centre.
The satellite's true cluster-centric distance, $R_{\rm sat}^{\rm true}$, is related to \redmapper assigned distance $R_{\rm sat_0}$ by the cosine law as, 
\begin{equation}
   R_{\rm sat}^{\rm true 2}=R^2_{\rm sat_0} + R^2_{\rm off} -2R_{\rm sat}R_{\rm off}\cos(\theta)\,. 
\end{equation}
Where $R_{\rm off}$ is the off-centred distance from the BCG. Practically, we carry out the integral over $R_{\rm off}$ until $10$ \mpch, and we have checked that the integral is well converged by varying this value by about 30 percent. This complete model to compute $\Delta \Sigma^{\rm DM}_{\rm clu}(R)$ is extensively described in Sec 4.2 of \citetalias{Kumar2022}; we suggest the reader refer to it for details. 

Their baryonic mass dominates the lensing signal around satellites in the innermost regions (e.g.  $R \lesssim 30$ kpc). As we move farther from the satellite centre, the subhalo mass of the satellites starts to dominate the lensing signal. The signal contribution due to the host cluster $\Delta\Sigma_{\rm clu}^{\rm DM}$ is close to zero at projected distances close to the satellite, as the difference between surface mass density within $R$ is not very different from the average density at $R$ due to the cluster halo. For distances corresponding to the central-satellite separations, i.e. $R~\sim ~R_{\rm sat}$, the $\Delta\Sigma(R)$ exceeds $\Delta\Sigma(<R)$, and the excess surface mass density becomes negative. The $\Delta\Sigma_{\rm clu}^{\rm DM}$ attains its negative peak at $R$ = $R_{\rm sat}$, a positive hump at $R$ = $2R_{\rm sat}$, and starts declining beyond this distance (See Fig.6 in \citetalias{Kumar2022}).     

In observations, we measure the weak lensing signal around satellite galaxies located at different cluster-centric distances and stack the signal for satellites falling within a particular radial bin to compute the average signal. This smoothens out the negative peak of $\Delta\Sigma_{\rm clu}^{\rm DM}$ and broadens the hump in a positive direction. Similarly, while modelling, we average out the lensing signal for all possible satellite distances in that particular radial bin as,   
\begin{equation}
\overline{\Delta \Sigma}(R|\sigma_{\rm mc}) = \int  P(R_{\rm sat_0})  \Delta \Sigma(R|R_{\rm sat_0},\sigma_{\rm mc})  dR_{\rm sat}.
\label{Pub3_eq:esd_final}
\end{equation}
where $P(R_{\rm sat_0})$ denotes the probability distribution of the cluster-centric distances of the satellites in a given bin, and $\Delta\Sigma$ is given by
\begin{equation}
\begin{split}
\Delta \Sigma(R|R_{\rm sat},\sigma_{\rm mc}) &=   \Delta\Sigma^{\rm bary}_{\rm sat}(R) + \Delta \Sigma^{\rm DM}_{\rm sat}(R)  \\&  \qquad\qquad\qquad\qquad+ \Delta\Sigma^{\rm DM}_{\rm clu}(R|R_{\rm sat_0},\sigma_{\rm mc})
\end{split}
\end{equation}

Finally, we point out that we expect some of the subhalos to have their profiles truncated due to the tidal forces of the galaxy cluster halo. However, the exact tidal radius does not depend upon the current position of the satellite but the closest pericentric passage of that satellite in 3-D. In addition, given that our satellites are selected in projection, we do not include any tidal effects to avoid using a single truncation radius for all satellites. The effect of tidal truncation on the weak lensing signal of a single satellite subhalo causes the signal to fall off as $1/R^2$ after the tidal truncation radius. 
\begin{figure*}
    \begin{subfigure}[b]{\textwidth}
        \hspace{0.8cm}\includegraphics[width=0.5\columnwidth]{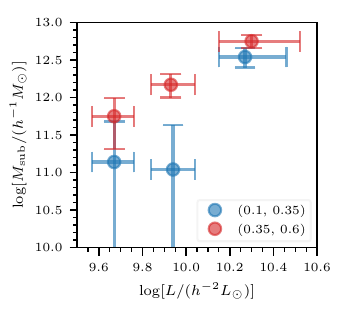}	
        \hspace{-0.5cm}
        \raisebox{0.2cm}{        \includegraphics[width=0.43\columnwidth]{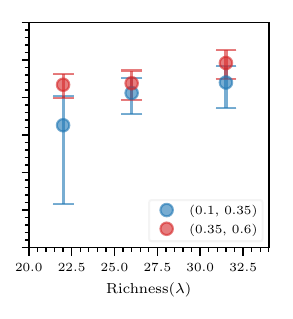}}            
     \end{subfigure}
\caption{The figure depicts the relationship of the subhalo mass with the satellite's luminosity and the host cluster's richness. The different colour schemes, i.e. blue and red, indicate the analysis for satellites from the inner and outer cluster centric radial bins of $R_{\rm sat_0}$. The analyses suggest that satellites evolving at faraway distances from the cluster centre are more massive compared to those evolving at closer separations. 
}	
\label{Pub3_fig:mass_luminosity_and_richness}    
\end{figure*}

In summary, our model has six parameters, i.e. $\rm  M_{clu}$, $\rm c_{clu}$, $\rm M_{sat}$, $\rm c_{sat}$, $\rm M_{bary}$, $\sigma_{\rm mc}$ signifying mass \& concentration of the host cluster and subhalo, average baryonic mass of the satellite and the width of the 2-D Rayleigh probability distribution used for accounting the miscentering of the central galaxy in our clusters, respectively.

\subsection{Fitting model to observations}
We perform a Bayesian analysis to infer the posterior distribution of our model parameter given the measurements of the lensing signal. In our six-parameter model discussed above, the concentration of the parent halo and subhalo is a free parameter. While performing the Bayesian analysis, we build upon the publicly available software AUM \footnotemark \footnotetext{https://github.com/surhudm/aum} for our modelling. We use the concentration-mass relation from \cite{Macci_2007} as our default model. We have also run chains that leave the concentration as a free parameter and we find that it does not affect our mass constraints significantly. To break the degeneracy between the subhalo mass of the satellite and its stellar mass while fitting the lensing signal, we use the stellar masses for \redmapper satellites as determined from the modelling of the HSC fluxes. In particular, we use the stellar masses from {\sc mizuki} photometric template fitting code used on HSC data by cross-matching the \redmapper satellites. Our model parameter $M_{\rm bary}$ takes the weighted mean of the HSC stellar mass. We use uninformative flat priors spanning wide sample space for each of the free parameters in our model, which are listed in Table \ref{Pub3_table:priors}. 

The posterior distribution of the parameters ($\Omega$) given the observed data vector (${\mathscr{D}}$) is given by the Bayes theorem as,
\begin{equation}
    P(\Theta|\mathscr{D}) \propto P({\mathscr{D}}|\Theta) P(\Theta)\,.
\end{equation}
Here $P(\mathscr{D}|{\Theta})$ represent the likelihood, and $P(\Theta)$ refers to the prior probability distribution of the model parameters. We assume the likelihood to be a Gaussian such that 
\begin{equation}
    \centering
    P({\mathscr{D}}|\Theta) \propto {\rm exp} \Big[-\frac{\chi^2(\Theta)}{2} \Big] P(\Theta). 
\end{equation}
In this representation 
\begin{equation}
\chi^2(\Theta) = [\Delta\Sigma_{\rm mod}-\Delta\Sigma_{\rm obs}]^T C^{-1} [\Delta\Sigma_{\rm mod}-\Delta\Sigma_{\rm obs}],
\label{Pub3_eq:likelihood}
\end{equation}
$\Delta\Sigma_{\rm mod}$ and $\Delta\Sigma_{\rm obs}$ denote the vector of model predictions and measured lensing signal, respectively. Whereas $C^{-1}$  represents the inverse of the smoothed covariance matrix obtained by the jackknife technique.

We use the publicly available package emcee\footnotemark \footnotetext{https://github.com/dfm/emcee} \citep{mcmc_hogg_2013} to perform a Monte Carlo Markov Chain(MCMC) analysis. We use 100 walkers running over 20000 steps, along with  500 burn-in steps, which we discard, to allow the chain to reach an equilibrium state.

\section{Results and Discussion}
\label{Pub3_sec:results_and_discussion}
\begin{figure*}
    \includegraphics[width=0.9\textwidth]{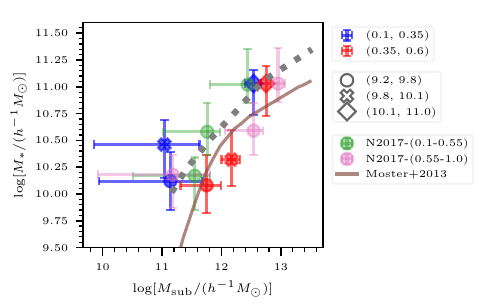}			
    \caption{The figure shows the comparison of the stellar to subhalo mass relations obtained from the luminosity-based binning scheme used in our analysis, with results from the literature. The blue and red colour schemes in the figure represent inner and outer radial bins of $R_{\rm sat_0}$, while the different markers (for the blue and red colour data only) represent luminosity cuts used in our analysis.}
\label{Pub3_fig:SHMR_inner_vs_outer}     
\end{figure*}

\begin{table*}
    \centering
    \begin{tabularx}{0.98\textwidth}{
    >{\centering\arraybackslash}X
    >{\centering\arraybackslash}X
    >{\centering\arraybackslash}X
    >{\centering\arraybackslash}X 
    >{\centering\arraybackslash}X 
    >{\centering\arraybackslash}X 
    }
    \hline\hline \\[-1.5ex]
	\multicolumn{1}{l}{} & \multicolumn{1}{c}{\textbf{$R_{\rm sat_0}$}} & \multicolumn{1}{c}{$\log M_{\rm clu}$} & \multicolumn{1}{c}{$\log M_{\rm sub}$} & \multicolumn{1}{c}{$\log M_{\rm bary}$}& \multicolumn{1}{c}{$\sigma_{\rm mc}$} \\[0.75ex]
    \hline\\[-2.5ex]
    \multicolumn{2}{l}{Luminosity/log($h^{-2}L_{\odot}$)}&   \multicolumn{3}{c}{} \\    
     \multirow{2}{*}{(9.2-9.8]}  & (0.1,0.35] & $14.36^{+0.03}_{-0.04}$ &$ 11.14^{+0.54}_{-1.20} $ & $10.12^{+0.27}_{-0.27}$ & $0.22^{+0.44}_{-0.16}$ \\[1.5ex]
     & (0.35,0.6]& $14.36^{+0.03}_{-0.03}$  & $11.75^{+0.24}_{-0.44}$ & $10.08^{+0.28}_{-0.26}$ & $0.32^{+0.41}_{-0.23}$  \\[2.5ex]
     \multirow{2}{*}{(9.8-10.1]}  & (0.1,0.35] & $14.35^{+0.03}_{-0.04}$ &$ 11.04^{+0.59}_{-1.19} $ & $10.46^{+0.23}_{-0.31}$ & $0.34^{+0.44}_{-0.26}$ \\[1.5ex]
     & (0.35,0.6]& $14.36^{+0.03}_{-0.03}$  & $12.17^{+0.14}_{-0.17}$ & $10.32^{+0.28}_{-0.25}$& $0.35^{+0.42}_{-0.25}$  \\[2.5ex]
     \multirow{2}{*}{(10.1-11.0]}  & (0.1,0.35] & $14.32^{+0.04}_{-0.04}$ &$ 12.54^{+0.12}_{-0.14} $ & $11.03^{+0.13}_{-0.29}$& $0.48^{+0.35}_{-0.33}$ \\[1.5ex]
     & (0.35,0.6]& $14.37^{+0.03}_{-0.03}$  & $12.75^{+0.07}_{-0.08}$ &$11.03^{+0.16}_{-0.30}$ & $0.28^{+0.43}_{-0.21}$  \\[2.5ex]
     \hdashline
     \\[-1.2ex]
     \multirow{1}{*}{(9.2-9.8]}  & (0.1,0.6] & $14.37^{+0.03}_{-0.03}$ &$ 11.58^{+0.21}_{-0.34} $ & $10.10^{+0.27}_{-0.27}$& $0.22^{+0.43}_{-0.16}$ \\[1.5ex]
     \multirow{1}{*}{(9.8-10.1]}  & (0.1,0.6] & $14.36^{+0.03}_{-0.03}$ &$ 11.80^{+0.16}_{-0.21} $ & $10.38^{+0.26}_{-0.28}$& $0.23^{+0.46}_{-0.17}$ \\[1.5ex]
     \multirow{1}{*}{(10.1-11.0]}  & (0.1,0.6] & $14.35^{+0.03}_{-0.03}$ &$ 12.57^{+0.08}_{-0.08} $ & $11.15^{+0.06}_{-0.15}$& $0.30^{+0.45}_{-0.22}$ \\[1.5ex]
     
     \hline\\[-2.5ex]
     \multicolumn{2}{l}{Richness ($\lambda$)}& \multicolumn{3}{c}{} \\[1.ex]    
     \multirow{2}{*}{(20.0-24.0]}  & (0.1,0.35] & $14.08^{+0.04}_{-0.05}$ &$ 11.63^{+0.39}_{-1.05} $ & $10.49^{+0.27}_{-0.28}$& $0.50^{+0.33}_{-0.34}$ \\[1.5ex]
     & (0.35,0.6]& $14.15^{+0.05}_{-0.05}$  & $12.17^{+0.14}_{-0.18}$ & $10.65^{+0.20}_{-0.32}$ & $0.47^{+0.35}_{-0.32}$  \\[2.5ex]
     \multirow{2}{*}{(24.0-28.0]}  & (0.1,0.35] & $14.07^{+0.04}_{-0.04}$ &$ 12.06^{+0.20}_{-0.28} $ & $10.47^{+0.27}_{-0.28}$& $0.41^{+0.41}_{-0.30}$ \\[1.5ex]
     & (0.35,0.6]& $14.08^{+0.04}_{-0.04}$  & $12.19^{+0.17}_{-0.22}$ & $10.59^{+0.24}_{-0.30}$ & $0.45^{+0.38}_{-0.32}$  \\[2.5ex]
     \multirow{2}{*}{(28.0-35.0]}  & (0.1,0.35] & $14.28^{+0.04}_{-0.04}$ &$ 12.20^{+0.22}_{-0.34} $ & $10.44^{+0.28}_{-0.25}$ & $0.37^{+0.41}_{-0.27}$ \\[1.5ex]
     & (0.35,0.6]& $14.35^{+0.04}_{-0.04}$  & $12.46^{+0.17}_{-0.21}$ & $10.49^{+0.27}_{-0.27}$ & $0.35^{+0.43}_{-0.26}$  \\[2.5ex]
     \hline\\[-2.5ex]
    \end{tabularx}
    \caption{\textit{Parameter Estimation}: The table lists the median values along with errors based on the 16 and 84 percentile from the posterior distributions with MCMC sampling for different fit parameters used in the analysis. The upper table quantifies the posterior estimations from binning schemes based on luminosity combined with cluster-centric radial separations $R_{\rm sat_0}$. Whereas the lower table denotes the results from a similar analysis performed using a richness-based binning scheme instead of luminosity.} 
\label{Pub3_table:infered_parameters}
\end{table*}

\begin{figure*}	\includegraphics[width=0.96\textwidth]{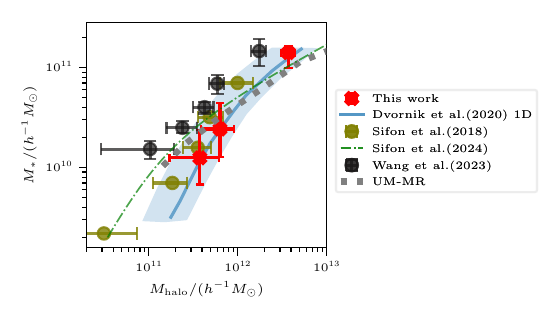}
	\caption{The figure shows a relationship between stellar and halo masses for satellite galaxies. The red colour in the figure shows sSHMRs results for our satellite population binned based on their luminosity. Whereas the orange and blue lines show the sSHMRs results using weak gravitational lensing around GAMA satellites by \citealt{Dvornik2020}. The olive colour data represents lensing results for satellites from low redshift clusters by \citealt{Sifon2018}. The green dashed-dotted line here refers to the best-fit double power law model of sSHMR from EAGLE simulations by \citealt{Sifon2024}.  Lastly, the black colour data depicts the results for subhalo masses for redMapper satellites by \citealt{Wang2024}.}
    \label{Pub3_fig:SHMR}
\end{figure*} 

We define samples of \redmapper satellite galaxies binned by their luminosity as well as the richness of the cluster in which they reside based on sample selection cuts listed in Table \ref{Pub3_table:satellies_data}. We further bin each of these samples of satellite galaxies by their cluster-centric radial separation in order to understand the impact of their environments on their subhalo masses. We measure the lensing signal for all these subsamples of \redmapper satellites as a function of distance from their centre $R$, following the procedure described in Sec.~\ref{Pub3_sec:weak_lensing_theory}. In order to account for any residual large-scale systematic bias in shape measurements, we remove the signal around the random points from the lensing signal. 

We show the measured weak lensing signal as green points with errors for satellite galaxies binned in luminosity in various columns of subpanels of Fig.~\ref{Pub3_fig:signal_luminosity_bins}. The top row subfigures correspond to satellite galaxies in different luminosity bins but evolving at the same cluster centric separation, i.e. $R_{\rm sat_0}$ $\in$ (0.1,0.35] \mpch, whereas the bottom row shows the similar analysis for satellites having separation of $R_{\rm sat_0}$ $\in$ (0.35,0.6] \mpch. The weak lensing signal due to the subhalo of satellite galaxies increases as a function of the luminosity of galaxies, signifying the larger subhalo masses of these satellite galaxies, presumably due to the larger halo masses in which these galaxies form. We also see a decrease in the weak lensing signal in the inner radial bins for galaxies of the same luminosity. The signal-to-noise ratio (SNR) for the lensing signal in each of the luminosity bins, i.e.(9.2,9.8], (9.8,10.1], (10.1,11.0] is [24.1, 24.5, 24.4 ],[23.3, 25.2, 28.05] corresponding to $R_{\rm sat_0}$ $\in$ (0.1,0.35] and (0.35,0.6], respectively. We will also present results for satellite galaxies in luminosity bins but combining all the cluster-centric distance bins.

In Fig.~\ref{Pub3_fig:signal_richness_bins}, we show the weak lensing signal of satellite galaxies separated by the same richness of their host cluster in each column and the same cluster centric distance of the satellites in a given row. In contrast to the trends with the luminosity of satellite galaxies, we see very little change in the weak lensing signal with the richness of the host galaxy clusters, although we do observe a consistent trend of a weaker signal in the radial bin closer to the cluster centre. The SNR for the weak lensing signal when binned by cluster's richness, i.e. (20.0,24.0],(24.0,28.0], (28.0,35.0] in $R_{\rm sat_0}$ $\in$ (0.1,0.35], (0.35,0.6] bins are [13.0, 19.15, 18.8],[12.2, 12.8, 16.4], respectively. 

The observed signals in each of the panels of both figures mimic our expectations for the lensing signal around satellites \citepalias{Kumar2022}. The signals in the innermost radial distances from the satellite galaxies are dominated by the subhalo mass, while the outer radial bins are related to the host halo of the galaxy cluster. We fit the measured signal with our theoretically motivated model described in Sec.~\ref{Pub3_sec:modelling}. Our model has free parameters for the subhalo mass of the satellite and the mass of its host cluster. We construct a joint residual vector $D (= \Delta\Sigma_{\rm Obs} - \Delta\Sigma_{\rm Model}  )$ for every parameter set in the MCMC and use it in the Eq. \ref{Pub3_eq:likelihood} to perform a Bayesian analysis.

The solid green lines, along with the shaded region in different panels of Fig.~\ref{Pub3_fig:signal_luminosity_bins}, represent the median model fit to the lensing signal and the 68 percent credible intervals from the MCMC samples. The goodness of fit, as indicated by the $\chi^2/{\rm dof}$, and the selection cuts on satellites corresponding to each bin are mentioned in each subfigure. We have listed the inferred parameters, along with their 68 percent credible intervals from the posterior distribution in Table \ref{Pub3_table:infered_parameters}. The $\chi^2/{\rm dof}$ for the signal fit using MCMC posterior estimations for most of the bins is well within the acceptable range. Although, in some of the cases, the reduced  $\overline{\chi^2}$ is a little high, we note this is driven by large scales attributable to the mass of the host clusters and does not have appreciable effects on our conclusions. Regardless, the cluster halo masses that we obtain $M\sim 2\times10^{14}$\msunH are comparable to those reported in the literature \citep{Miyatake2016, More_2016, Simet2017, Murata2018}, despite the fact that our weak lensing signal effectively only probes the large scales.

In Fig.~\ref{Pub3_fig:Posterior_distribution}, we show the posterior distribution obtained from the MCMC sampling for our model parameters. The left and the right panels in the figure show the 2-D histograms for the analysis in one of the luminosity and richness-based binning schemes, respectively. These figures are representative of what we see for other bins also. We describe the summary statistics of posterior distributions for the model parameters in Table~\ref{Pub3_table:infered_parameters}. The subhalo mass of the satellite and their host halo masses are determined with an accuracy that varies from bin to bin depending upon the number statistics of satellites, as well as the amplitude of the signal. We observe the mass of the cluster halo has a minor degeneracy with the subhalo mass. We also find that the nuisance parameter corresponding to the off-centring of the BCG has very little degeneracy with the parameters of our interest since the weak lensing signal of the subhalo associated with the satellite galaxy is well separated from the signal from the main cluster halo. 

\subsection{Cluster-richness and luminosity dependence of subhalo mass of the satellites}
\label{Pub3_luminosity_mass}
The two panels of Fig.~\ref{Pub3_fig:mass_luminosity_and_richness} show the dependence of the inferred subhalo mass of the satellites from our weak lensing analysis on the luminosity of satellite galaxies and the host cluster's richness, respectively. The errorbars on $M_{\rm sub}$ in the figure represent the 1-$\sigma$ credible interval obtained from the posterior distribution of the corresponding model parameter, and the scatter in luminosity corresponds to the 16 and 84 percentile values of the satellite galaxy luminosity from the \redmapper catalogue. The blue and red coloured points with errorbars in each of the panels correspond to the results in the inner ( $R_{\rm sat_0}$ $\in (0.1,0.35]$) and outer $(0.35,0.6]$) cluster centric radial bins, respectively. 

The left panel of Fig.~\ref{Pub3_fig:mass_luminosity_and_richness} shows that more luminous satellite galaxies reside in more massive subhalos in the outermost cluster-centric distance bin. This is likely a consequence of the luminosity halo mass relation of galaxies before they became satellite galaxies. Deep into the cluster central regions, we see consistent lower subhalo masses compared to the outermost cluster-centric bin. However, we see a strong effect of stripping on the subhalo mass. The dependence of the subhalo mass on the luminosity for satellite galaxies with luminosity $L<10^{10}\lsunhH$ is almost wiped out by the loss of the subhalo mass. Galaxies near the cluster centres but more luminous than above, however, seem to survive this effect.

The right panel of Fig.~\ref{Pub3_fig:mass_luminosity_and_richness} depicts the relation between the inferred subhalo mass with the cluster richness. The inferred cluster masses $M_{\rm clu}$ increase with richness, and one could expect that the subhalo masses should also increase with richness. However, we observe very little difference in the average subhalo mass as a function of richness. This suggests that the very few bright satellites do not dominate the average subhalo masses. Given that the \redmapper satellites have a faint end cut, the more numerous fainter satellites drive the average subhalo masses to be similar. The trend with the inner and outer radial bins shows a consistent trend with respect to the binning by luminosity, and the subhalos in the inner cluster centric radial bins are less massive due to stripping. There is another competing effect at play, we expect that the more massive subhalos (corresponding to the more luminous galaxies), would sink to the centre. However, it seems tidal stripping dominates in the inner regions.

\subsection{Stellar to halo mass relations for satellites}
\label{Pub3_sec:shmr}

The SHMRs for satellites (hereafter sSHMRs) evolving in dense environments are expected to be different than that of field centrals, as satellites possess lower subhalo masses compared to their counterpart galaxies in the field environment \citep{Rodr_puebla2012, Rodr_puebla2013, Kumar2022}. The sSHMRs depend on the definition of subhalo mass being used, e.g. mass at accretion time $M^{\rm accr}_{\rm shalo}$ or mass at current epoch $M^{\rm obs}_{\rm shalo}$. However, the sSHMR is expected to increase gradually with the increase in satellite separation from the cluster centre. Motivated by this idea and to compare against existing results in the literature, we explore the subhalo mass evolution for satellites using luminosity-based selection cuts combined with cluster-centric radial separation $R_{\rm sat_0}$.

In Fig.~\ref{Pub3_fig:SHMR_inner_vs_outer}, we show a comparison of sSHMRs obtained from our analysis with a similar study by \citealt{Niemiec2017} using stellar mass-based binning scheme combined with radial separation from cluster centre for \redmapper satellites itself. \citealt{Niemiec2017} use different radial selection cuts, i.e. $R_{\rm sat_0}$ $\in$ (0.1,0.55),(0.55,1.0) \mpcH as inner and outer bin definitions and corresponding results are indicated with light-green and light-pink colours in the figure. In the same figure, the blue and red coloured points with errors represent the results for inner and outer radial bins of $R_{\rm sat_0}$ in our analysis, while the different marker types for these colours represent the different luminosity cuts in our analysis. The errorbars on the subhalo mass estimate represent the 16 and 84 percentile credible limits from the posterior chains. The figure shows our results are in qualitative agreement with \citealt{Niemiec2017} given the errors. The results from both analyses conform with the expectation that the present day sSHMRs can be affected by the distance of the satellite from the cluster centre; the subhalo to stellar mass ratio increases for satellites which are at present farther from the cluster centre. For the farthest distances from the cluster centre, the sSHMRs approach the theoretical prediction for isolated central galaxies from the abundance matching method. We show this abundance matching result from \citep{Moster2013} with the brown coloured solid line. Assuming the stellar mass of the satellite is not much affected after accretion, our results are in accordance with expectations from tidal stripping of the subhalo mass. 

We also compare our results against sSHMRs from the mock-\redmapper constructed on the publicly available UniverseMachine galaxy and halo catalogues. UniverseMachine (hereafter UM) is a model of galaxy formation and evolution that was applied to subhalo merger trees from the Bolshoi-Plank N-body simulation \citep{Rodr_puebla2016} and was calibrated using various observations, e.g. galaxy abundance, the cosmic star formation rate, as well as the auto and cross-correlation of star-forming and quenched galaxies star formation rate, at low to high redshifts \citep{Behroozi_2019}. We run a \redmapper like cluster finding algorithm on the UM catalogue to construct a mock-\redmapper (hereafter MR) cluster and member catalogue out of the UM \citep[see][ for more details]{Sunayama2019, Kumar2024}. This allows us to also account for any interlopers that may affect the sSHMR from observations. We plot the sSHMRs obtained for satellite galaxies in MR in the same cluster-centric distance bins as in our analysis, with the dotted grey line in Fig. \ref{Pub3_fig:SHMR_inner_vs_outer}. We use the weak lensing baryonic mass of the satellites in the figure, and the results from the UM-MR are in good agreement with our observed masses. Although the exact prescription for how the subhalos and their masses evolve in the UM with respect to orphan galaxies is an uncertainty \citep{Kumar2024}, the subhalo mass loss does reproduce the overall sSHMR from observations well.

Till now, we explored the effects of luminosity or richness-based binning schemes on subhalo masses of satellites evolving at different separations from the cluster centre. Next, we compare the sSHMRs obtained from the luminosity-based binning schemes for our satellites with results on sSHMRS from the literature in Fig. \ref{Pub3_fig:SHMR}. For this purpose, we combined the cluster-centric radial bins into one single bin, measured the weak lensing signal, and inferred the subhalo masses of our satellite galaxies by following the same luminosity-based selection cuts. The red colour data points with errors in the figure show the results of our analysis. The blue solid line and the corresponding shaded region represent the weak lensing results for the sSHMR for GAMA satellites using data from the KiDS survey by \citet{Dvornik2020}. The olive colour points with errors in the figure represent the sSHMRs using weak lensing measurements around spectroscopically confirmed + red sequence satellite galaxies from low redshift MENeaCS clusters in CHFT survey, by \citet{Sifon2018}. Whereas the dashed green line in the figure depicts the best fit to the double power law model (see their Eq.1) of SHMRs for satellites in clusters with mass $M_{\rm 200m}> 10^{13}$ from EAGLE simulations, by \citet{Sifon2024}. Finally, the black coloured points with errors show the lensing results for sSHMR from \citet{Wang2024}, for redMapper satellites using DECaLS DR8 data. Our results are in overall qualitative agreement with the satellite mass study of \citet{Sifon2018, Sifon2024} and \citet{Dvornik2020}. 

Finally, we mention that although we see qualitative consistency with \citet{Wang2024} who also use the satellites from the \redmapper cluster catalogue, we do observe some quantitative differences. The subhalo masses from \citet{Wang2024} are systematically lower than the subhalo masses in our study. We speculate that this might be a result of the use of a definition of subhalo mass, which corresponds to the mass enclosed in a radius where the subhalo density equals that of the main cluster halo. The rationale behind the use of this definition was to compare the subhalo masses with those from subfind, an algorithm which uses similar criteria to select subhalos \citep{Springel2001}. However, one needs to estimate how big these effects are, or if these differences are related to differences in our satellite galaxy sample selection or the uncertainties in the source photometric redshifts used for their weak lensing analyses.

\section{Summary}
\label{Pub3_sec:summary}
In this paper, we make use of satellite galaxies defined in the \redmapper cluster catalogue and lying within the redshift range of $0.1\leq z\leq 0.33$, to understand the relation between galaxy observables such as luminosity and stellar mass with their subhalo masses, using the weak gravitational lensing of source galaxies from the Subaru HSC survey Y3 release. In particular, we infer the luminosity-halo mass relations for satellites (sLHMR) and the stellar-to-halo mass relation (sSHMR) for satellite galaxies. Our results and key findings can be summarized as follows: 

\begin{itemize}

    \item  We binned our satellite galaxies based on their luminosity and the richness of the host cluster combined with the cluster-centric projected separations of the satellites to understand the impact of the host environment on their evolution. The weak lensing signals were measured with a good signal-to-noise ratio in each case (SNR $>10$).

    \item Our physically motivated model for satellite galaxies is able to explain the observed weak lensing signal around satellite galaxies reasonably well. Our model accounts for the mass profile associated with the subhalo of the satellites, as well as that around the galaxy cluster to which the satellite belongs.

    \item We find that the subhalo masses of satellites that evolve closer to the cluster centre ($R_{\rm sat,0}<0.35$ \mpch) are lower than satellites having the same luminosity but evolving at farther separations in the outer regions of galaxy clusters ($0.35< R_{\rm sat,0}/{\rm h^{-1}Mpc} <0.6$). This is consistent with expectations that the strength of tidal interactions experienced by satellite galaxies increases with decreasing distance from the cluster centre. The luminosity-halo mass relation for satellite (sLHMR) of satellite galaxies tends towards the central galaxy relation in the outer regions, with more luminous satellite galaxies that reside in more massive subhalos, similar to the luminosity-halo mass relation of galaxies before they became satellites. 
.
    \item In the inner cluster-centric distance bins, we find that the sLHMR of satellite galaxies shows no evidence of increase with luminosity for galaxies with $L<10^{10}\lsunhh$. This implies that the mass stripping of subhalos dominates for low-luminosity galaxies evolving closer to the cluster centre. Highly luminous galaxies seem to survive the effects of tidal stripping comparatively better.

    \item Clusters with high richness are expected to host more luminous and heavier satellites. However, the average subhalo masses do not show much evidence of a dependence on the richness of the cluster. This is likely due to the fact that the more numerous fainter galaxies dominate the statistics.
    
    \item  We do not see clear evidence of subhalo segregation by mass either, suggesting that even though more massive subhalos corresponding to the luminous galaxies sink to the centre, these massive subhalos also get stripped of the mass around them, removing any trend of increasing subhalo masses with a decrease in the distance from the cluster centre.
    
    \item We also show that the sSHMRs can be affected by the distance of the satellites from the cluster centre, and the subhalo mass to stellar mass ratio for satellites increases as the separation of the satellite increases from the cluster centre. We show that our results are consistent with other results in the literature.
    
    \item We also compare our results of the inferred sSHMR for satellites in the mock \redmapper cluster catalogue constructed based on the UniverseMachine model of simulated galaxies and find good agreement with our observational results.
\end{itemize}

With a simple model to fit the observed lensing signal around satellites, we were able to measure subhalo and cluster masses with an accuracy that depends upon the strength of the signal and the number of satellite galaxies in our samples. Complicated theoretical models accounting for dark matter nature, two halo terms may not be needed right now, given the statistical precision available with the available data. However, observations from the upcoming Legacy Survey of Space and Time (LSST) with the Vera Rubin observatory will pin down the statistical uncertainties by a further factor of 5 to 6, and will help constrain more physical signatures of environmental evolution, such as the tidal truncation radius of the subhalos.

\section{Data Availability}
The satellite galaxy sample used in this analysis is constructed from publicly available \redmapper cluster/member catalogue with redshift and P$_{\rm cen}$ as described in  Sec:\ref{Pub3_sec:redmapper}. The 2D - histograms of the posterior distribution for all the bins used in the analysis are available at \@ \url{https://github.com/ratewalamit/mass_luminosity_DAV.git}.

\section*{Acknowledgements}

The author would like to thank Navin Chaurasiya, Divya Rana, Susmita Adhikari, Arka Banerjee, Preetish Mishra, Priyanka Gawade, and Joy Bhattacharyya for the useful discussions/comments/suggestions/inputs throughout this work and on the draft version of this manuscript. The authors acknowledge the use of Pegasus, the high-performance computing (HPC) facility of IUCAA.
\bibliographystyle{mnras}
\bibliography{bibliography} 

\begin{thebibliography}{}
\makeatletter
\relax
\def\mn@urlcharsother{\let\do\@makeother \do\$\do\&\do\#\do\^\do\_\do\%\do\~}
\def\mn@doi{\begingroup\mn@urlcharsother \@ifnextchar [ {\mn@doi@} {\mn@doi@[]}}
\def\mn@doi@[#1]#2{\def\@tempa{#1}\ifx\@tempa\@empty \href {http://dx.doi.org/#2} {doi:#2}\else \href {http://dx.doi.org/#2} {#1}\fi \endgroup}
\def\mn@eprint#1#2{\mn@eprint@#1:#2::\@nil}
\def\mn@eprint@arXiv#1{\href {http://arxiv.org/abs/#1} {{\tt arXiv:#1}}}
\def\mn@eprint@dblp#1{\href {http://dblp.uni-trier.de/rec/bibtex/#1.xml} {dblp:#1}}
\def\mn@eprint@#1:#2:#3:#4\@nil{\def\@tempa {#1}\def\@tempb {#2}\def\@tempc {#3}\ifx \@tempc \@empty \let \@tempc \@tempb \let \@tempb \@tempa \fi \ifx \@tempb \@empty \def\@tempb {arXiv}\fi \@ifundefined {mn@eprint@\@tempb}{\@tempb:\@tempc}{\expandafter \expandafter \csname mn@eprint@\@tempb\endcsname \expandafter{\@tempc}}}

\bibitem[\protect\citeauthoryear{Aihara et~al.,}{Aihara et~al.}{2011}]{Aihara2011}
Aihara H.,  et~al., 2011, \mn@doi [\apjs] {10.1088/0067-0049/193/2/29}, \href {https://ui.adsabs.harvard.edu/abs/2011ApJS..193...29A} {193, 29}

\bibitem[\protect\citeauthoryear{Aihara et~al.,}{Aihara et~al.}{2017}]{aihara2017}
Aihara H.,  et~al., 2017, \mn@doi [Publications of the Astronomical Society of Japan] {10.1093/pasj/psx066}, 70

\bibitem[\protect\citeauthoryear{Aihara et~al.,}{Aihara et~al.}{2018}]{Aihara2018}
Aihara H.,  et~al., 2018, \mn@doi [\pasj] {10.1093/pasj/psx081}, \href {https://ui.adsabs.harvard.edu/abs/2018PASJ...70S...8A} {70, S8}

\bibitem[\protect\citeauthoryear{Aihara et~al.,}{Aihara et~al.}{2019}]{aihara2019}
Aihara H.,  et~al., 2019, \mn@doi [Publications of the Astronomical Society of Japan] {10.1093/pasj/psz103}, 71

\bibitem[\protect\citeauthoryear{{Bartelmann} \& {Schneider}}{{Bartelmann} \& {Schneider}}{2001}]{Bartelmann_and_Schneider_2001}
{Bartelmann} M.,  {Schneider} P.,  2001, \mn@doi [\physrep] {10.1016/S0370-1573(00)00082-X}, \href {https://ui.adsabs.harvard.edu/abs/2001PhR...340..291B} {340, 291}

\bibitem[\protect\citeauthoryear{{Behroozi}, {Wechsler}, {Hearin}  \& {Conroy}}{{Behroozi} et~al.}{2019}]{Behroozi_2019}
{Behroozi} P.,  {Wechsler} R.~H.,  {Hearin} A.~P.,   {Conroy} C.,  2019, \mn@doi [\mnras] {10.1093/mnras/stz1182}, \href {https://ui.adsabs.harvard.edu/abs/2019MNRAS.488.3143B} {488, 3143}

\bibitem[\protect\citeauthoryear{{Bernstein} \& {Jarvis}}{{Bernstein} \& {Jarvis}}{2002}]{Bernstein_Jarvis_2002}
{Bernstein} G.~M.,  {Jarvis} M.,  2002, \mn@doi [\aj] {10.1086/338085}, \href {https://ui.adsabs.harvard.edu/abs/2002AJ....123..583B} {123, 583}

\bibitem[\protect\citeauthoryear{{Bhattacharyya}, {Adhikari}, {Banerjee}, {More}, {Kumar}, {Nadler}  \& {Chatterjee}}{{Bhattacharyya} et~al.}{2022}]{Bhattacharyya_2021}
{Bhattacharyya} S.,  {Adhikari} S.,  {Banerjee} A.,  {More} S.,  {Kumar} A.,  {Nadler} E.~O.,   {Chatterjee} S.,  2022, \mn@doi [\apj] {10.3847/1538-4357/ac68e9}, \href {https://ui.adsabs.harvard.edu/abs/2022ApJ...932...30B} {932, 30}

\bibitem[\protect\citeauthoryear{{Binney} \& {Tremaine}}{{Binney} \& {Tremaine}}{2008}]{Binney2008}
{Binney} J.,  {Tremaine} S.,  2008, {Galactic Dynamics: Second Edition}.
Princeton University Press

\bibitem[\protect\citeauthoryear{{Blanton} \& {Roweis}}{{Blanton} \& {Roweis}}{2007}]{Blanton2007}
{Blanton} M.~R.,  {Roweis} S.,  2007, \mn@doi [\aj] {10.1086/510127}, \href {https://ui.adsabs.harvard.edu/abs/2007AJ....133..734B} {133, 734}

\bibitem[\protect\citeauthoryear{{Bose}, {Eisenstein}, {Hernquist}, {Pillepich}, {Nelson}, {Marinacci}, {Springel}  \& {Vogelsberger}}{{Bose} et~al.}{2019}]{Bose2019}
{Bose} S.,  {Eisenstein} D.~J.,  {Hernquist} L.,  {Pillepich} A.,  {Nelson} D.,  {Marinacci} F.,  {Springel} V.,   {Vogelsberger} M.,  2019, \mn@doi [\mnras] {10.1093/mnras/stz2546}, \href {https://ui.adsabs.harvard.edu/abs/2019MNRAS.490.5693B} {490, 5693}

\bibitem[\protect\citeauthoryear{{Brainerd}, {Blandford}  \& {Smail}}{{Brainerd} et~al.}{1996}]{Brainerd1996}
{Brainerd} T.~G.,  {Blandford} R.~D.,   {Smail} I.,  1996, \mn@doi [\apj] {10.1086/177537}, \href {https://ui.adsabs.harvard.edu/abs/1996ApJ...466..623B} {466, 623}

\bibitem[\protect\citeauthoryear{{Chang}, {Macci{\`o}}  \& {Kang}}{{Chang} et~al.}{2013}]{Chang2013}
{Chang} J.,  {Macci{\`o}} A.~V.,   {Kang} X.,  2013, \mn@doi [\mnras] {10.1093/mnras/stt434}, \href {https://ui.adsabs.harvard.edu/abs/2013MNRAS.431.3533C} {431, 3533}

\bibitem[\protect\citeauthoryear{Coupon et~al.,}{Coupon et~al.}{2015}]{Coupon2015}
Coupon J.,  et~al., 2015, \mn@doi [Monthly Notices of the Royal Astronomical Society] {10.1093/mnras/stv276}, 449, 1352

\bibitem[\protect\citeauthoryear{{Dalal} et~al.,}{{Dalal} et~al.}{2023}]{Dalal2023}
{Dalal} R.,  et~al., 2023, \mn@doi [\prd] {10.1103/PhysRevD.108.123519}, \href {https://ui.adsabs.harvard.edu/abs/2023PhRvD.108l3519D} {108, 123519}

\bibitem[\protect\citeauthoryear{{\VAN{DeJong}{De}{de}}~Jong et~al.,}{{\VAN{DeJong}{De}{de}}~Jong et~al.}{2013}]{de_Jong2013}
{\VAN{DeJong}{De}{de}}~Jong J.~T.~A.,  et~al., 2013, The Messenger, \href {https://ui.adsabs.harvard.edu/abs/2013Msngr.154...44D} {154, 44}

\bibitem[\protect\citeauthoryear{{\VAN{DeLucia}{De}{de}}~Lucia, Kauffmann, Springel, White, Lanzoni, Stoehr, Tormen  \& Yoshida}{{\VAN{DeLucia}{De}{de}}~Lucia et~al.}{2004}]{Lucia2004}
{\VAN{DeLucia}{De}{de}}~Lucia G.,  Kauffmann G.,  Springel V.,  White S. D.~M.,  Lanzoni B.,  Stoehr F.,  Tormen G.,   Yoshida N.,  2004, \mn@doi [Monthly Notices of the Royal Astronomical Society] {10.1111/j.1365-2966.2004.07372.x}, 348, 333

\bibitem[\protect\citeauthoryear{{Desmond}, {Mao}, {Wechsler}, {Crain}  \& {Schaye}}{{Desmond} et~al.}{2017}]{Desmond2017}
{Desmond} H.,  {Mao} Y.-Y.,  {Wechsler} R.~H.,  {Crain} R.~A.,   {Schaye} J.,  2017, \mn@doi [\mnras] {10.1093/mnrasl/slx093}, \href {https://ui.adsabs.harvard.edu/abs/2017MNRAS.471L..11D} {471, L11}

\bibitem[\protect\citeauthoryear{{Dey} et~al.,}{{Dey} et~al.}{2019}]{Dey2019}
{Dey} A.,  et~al., 2019, \mn@doi [\aj] {10.3847/1538-3881/ab089d}, \href {https://ui.adsabs.harvard.edu/abs/2019AJ....157..168D} {157, 168}

\bibitem[\protect\citeauthoryear{{Driver} et~al.,}{{Driver} et~al.}{2011}]{Driver2011}
{Driver} S.~P.,  et~al., 2011, \mn@doi [\mnras] {10.1111/j.1365-2966.2010.18188.x}, \href {https://ui.adsabs.harvard.edu/abs/2011MNRAS.413..971D} {413, 971}

\bibitem[\protect\citeauthoryear{{Dvornik} et~al.,}{{Dvornik} et~al.}{2020}]{Dvornik2020}
{Dvornik} A.,  et~al., 2020, \mn@doi [\aap] {10.1051/0004-6361/202038693}, \href {https://ui.adsabs.harvard.edu/abs/2020A&A...642A..83D} {642, A83}

\bibitem[\protect\citeauthoryear{{Flaugher} et~al.,}{{Flaugher} et~al.}{2015}]{Flaugher2015}
{Flaugher} B.,  et~al., 2015, \mn@doi [\aj] {10.1088/0004-6256/150/5/150}, \href {https://ui.adsabs.harvard.edu/abs/2015AJ....150..150F} {150, 150}

\bibitem[\protect\citeauthoryear{{Foreman-Mackey}, {Hogg}, {Lang}  \& {Goodman}}{{Foreman-Mackey} et~al.}{2013}]{mcmc_hogg_2013}
{Foreman-Mackey} D.,  {Hogg} D.~W.,  {Lang} D.,   {Goodman} J.,  2013, \mn@doi [\pasp] {10.1086/670067}, \href {https://ui.adsabs.harvard.edu/abs/2013PASP..125..306F} {125, 306}

\bibitem[\protect\citeauthoryear{{Frenk} \& {White}}{{Frenk} \& {White}}{2012}]{Frenk2012}
{Frenk} C.~S.,  {White} S.~D.~M.,  2012, \mn@doi [Annalen der Physik] {10.1002/andp.201200212}, \href {https://ui.adsabs.harvard.edu/abs/2012AnP...524..507F} {524, 507}

\bibitem[\protect\citeauthoryear{Furusawa et~al.,}{Furusawa et~al.}{2017}]{furusawa2017}
Furusawa H.,  et~al., 2017, \mn@doi [Publications of the Astronomical Society of Japan] {10.1093/pasj/psx079}, 70

\bibitem[\protect\citeauthoryear{{Gao}, {White}, {Jenkins}, {Stoehr}  \& {Springel}}{{Gao} et~al.}{2004}]{Gao2004}
{Gao} L.,  {White} S.~D.~M.,  {Jenkins} A.,  {Stoehr} F.,   {Springel} V.,  2004, \mn@doi [\mnras] {10.1111/j.1365-2966.2004.08360.x}, \href {https://ui.adsabs.harvard.edu/abs/2004MNRAS.355..819G} {355, 819}

\bibitem[\protect\citeauthoryear{Giocoli, Tormen  \& Van Den~Bosch}{Giocoli et~al.}{2008}]{Giocoli2008}
Giocoli C.,  Tormen G.,   Van Den~Bosch F.~C.,  2008, \mn@doi [Monthly Notices of the Royal Astronomical Society] {10.1111/j.1365-2966.2008.13182.x}, 386, 2135

\bibitem[\protect\citeauthoryear{{Girelli}, {Pozzetti}, {Bolzonella}, {Giocoli}, {Marulli}  \& {Baldi}}{{Girelli} et~al.}{2020}]{Girelli2020}
{Girelli} G.,  {Pozzetti} L.,  {Bolzonella} M.,  {Giocoli} C.,  {Marulli} F.,   {Baldi} M.,  2020, \mn@doi [\aap] {10.1051/0004-6361/201936329}, \href {https://ui.adsabs.harvard.edu/abs/2020A&A...634A.135G} {634, A135}

\bibitem[\protect\citeauthoryear{{Gunn} \& {Gott}}{{Gunn} \& {Gott}}{1972}]{Gunn1972}
{Gunn} J.~E.,  {Gott} J.~Richard I.,  1972, \mn@doi [\apj] {10.1086/151605}, \href {https://ui.adsabs.harvard.edu/abs/1972ApJ...176....1G} {176, 1}

\bibitem[\protect\citeauthoryear{{He} et~al.,}{{He} et~al.}{2022}]{He2022}
{He} Q.,  et~al., 2022, \mn@doi [\mnras] {10.1093/mnras/stac759}, \href {https://ui.adsabs.harvard.edu/abs/2022MNRAS.512.5862H} {512, 5862}

\bibitem[\protect\citeauthoryear{{He} et~al.,}{{He} et~al.}{2023}]{He2023}
{He} Q.,  et~al., 2023, \mn@doi [\mnras] {10.1093/mnras/stac2779}, \href {https://ui.adsabs.harvard.edu/abs/2023MNRAS.518..220H} {518, 220}

\bibitem[\protect\citeauthoryear{Heymans et~al.,}{Heymans et~al.}{2006}]{heymans2006}
Heymans C.,  et~al., 2006, Monthly Notices of the Royal Astronomical Society: Letters, 371, L60

\bibitem[\protect\citeauthoryear{{Heymans} et~al.,}{{Heymans} et~al.}{2012}]{Heymans2012}
{Heymans} C.,  et~al., 2012, \mn@doi [\mnras] {10.1111/j.1365-2966.2012.21952.x}, \href {https://ui.adsabs.harvard.edu/abs/2012MNRAS.427..146H} {427, 146}

\bibitem[\protect\citeauthoryear{{Hirata} \& {Seljak}}{{Hirata} \& {Seljak}}{2003}]{Hirata2003}
{Hirata} C.,  {Seljak} U.,  2003, \mn@doi [\mnras] {10.1046/j.1365-8711.2003.06683.x}, \href {https://ui.adsabs.harvard.edu/abs/2003MNRAS.343..459H} {343, 459}

\bibitem[\protect\citeauthoryear{{Hoekstra}, {Franx}, {Kuijken}  \& {van Dokkum}}{{Hoekstra} et~al.}{2002}]{Hoekstra2002}
{Hoekstra} H.,  {Franx} M.,  {Kuijken} K.,   {van Dokkum} P.~G.,  2002, \mn@doi [\mnras] {10.1046/j.1365-8711.2002.05479.x}, \href {https://ui.adsabs.harvard.edu/abs/2002MNRAS.333..911H} {333, 911}

\bibitem[\protect\citeauthoryear{Hoekstra, Hsieh, Yee, Lin  \& Gladders}{Hoekstra et~al.}{2005}]{Hoekstra2005}
Hoekstra H.,  Hsieh B.~C.,  Yee H. K.~C.,  Lin H.,   Gladders M.~D.,  2005, \mn@doi [The Astrophysical Journal] {10.1086/496913}, 635, 73

\bibitem[\protect\citeauthoryear{{Hsieh} \& {Yee}}{{Hsieh} \& {Yee}}{2014}]{Hseish2014}
{Hsieh} B.~C.,  {Yee} H.~K.~C.,  2014, \mn@doi [\apj] {10.1088/0004-637X/792/2/102}, \href {https://ui.adsabs.harvard.edu/abs/2014ApJ...792..102H} {792, 102}

\bibitem[\protect\citeauthoryear{{Johnston} et~al.,}{{Johnston} et~al.}{2007}]{Johnston2007}
{Johnston} D.~E.,  et~al., 2007, arXiv e-prints, \href {https://ui.adsabs.harvard.edu/abs/2007arXiv0709.1159J} {p. arXiv:0709.1159}

\bibitem[\protect\citeauthoryear{{Kawanomoto} et~al.,}{{Kawanomoto} et~al.}{2018}]{Kawanomoto2018}
{Kawanomoto} S.,  et~al., 2018, \mn@doi [\pasj] {10.1093/pasj/psy056}, \href {https://ui.adsabs.harvard.edu/abs/2018PASJ...70...66K} {70, 66}

\bibitem[\protect\citeauthoryear{{Kneib} \& {Natarajan}}{{Kneib} \& {Natarajan}}{2011}]{Kneib2011}
{Kneib} J.-P.,  {Natarajan} P.,  2011, \mn@doi [\aapr] {10.1007/s00159-011-0047-3}, \href {https://ui.adsabs.harvard.edu/abs/2011A&ARv..19...47K} {19, 47}

\bibitem[\protect\citeauthoryear{{Kneib}, {Ellis}, {Smail}, {Couch}  \& {Sharples}}{{Kneib} et~al.}{1996}]{Kneib1996}
{Kneib} J.~P.,  {Ellis} R.~S.,  {Smail} I.,  {Couch} W.~J.,   {Sharples} R.~M.,  1996, \mn@doi [\apj] {10.1086/177995}, \href {https://ui.adsabs.harvard.edu/abs/1996ApJ...471..643K} {471, 643}

\bibitem[\protect\citeauthoryear{Komiyama et~al.,}{Komiyama et~al.}{2017}]{komiyama2017}
Komiyama Y.,  et~al., 2017, \mn@doi [Publications of the Astronomical Society of Japan] {10.1093/pasj/psx069}, 70

\bibitem[\protect\citeauthoryear{{Koopmans}}{{Koopmans}}{2005}]{Koopmans2005}
{Koopmans} L.~V.~E.,  2005, \mn@doi [\mnras] {10.1111/j.1365-2966.2005.09523.x}, \href {https://ui.adsabs.harvard.edu/abs/2005MNRAS.363.1136K} {363, 1136}

\bibitem[\protect\citeauthoryear{{Kumar}, {More}  \& {Rana}}{{Kumar} et~al.}{2022}]{Kumar2022}
{Kumar} A.,  {More} S.,   {Rana} D.,  2022, \mn@doi [\mnras] {10.1093/mnras/stac2862}, \href {https://ui.adsabs.harvard.edu/abs/2022MNRAS.517.4389K} {517, 4389}

\bibitem[\protect\citeauthoryear{{Kumar}, {More}  \& {Sunayama}}{{Kumar} et~al.}{2024}]{Kumar2024}
{Kumar} A.,  {More} S.,   {Sunayama} T.,  2024, \mn@doi [\mnras] {10.1093/mnrasl/slae023}, \href {https://ui.adsabs.harvard.edu/abs/2024MNRAS.531L..20K} {531, L20}

\bibitem[\protect\citeauthoryear{{Leauthaud} et~al.,}{{Leauthaud} et~al.}{2012}]{Leauthaud2012}
{Leauthaud} A.,  et~al., 2012, \mn@doi [\apj] {10.1088/0004-637X/744/2/159}, \href {https://ui.adsabs.harvard.edu/abs/2012ApJ...744..159L} {744, 159}

\bibitem[\protect\citeauthoryear{{Li} et~al.,}{{Li} et~al.}{2014}]{Li2014}
{Li} R.,  et~al., 2014, \mn@doi [\mnras] {10.1093/mnras/stt2395}, \href {https://ui.adsabs.harvard.edu/abs/2014MNRAS.438.2864L} {438, 2864}

\bibitem[\protect\citeauthoryear{{Li} et~al.,}{{Li} et~al.}{2016}]{Li2016}
{Li} R.,  et~al., 2016, \mn@doi [\mnras] {10.1093/mnras/stw494}, \href {https://ui.adsabs.harvard.edu/abs/2016MNRAS.458.2573L} {458, 2573}

\bibitem[\protect\citeauthoryear{{Li} et~al.,}{{Li} et~al.}{2022}]{Li2022}
{Li} X.,  et~al., 2022, \mn@doi [\pasj] {10.1093/pasj/psac006}, \href {https://ui.adsabs.harvard.edu/abs/2022PASJ...74..421L} {74, 421}

\bibitem[\protect\citeauthoryear{{Li} et~al.,}{{Li} et~al.}{2023}]{Li2023}
{Li} X.,  et~al., 2023, \mn@doi [\prd] {10.1103/PhysRevD.108.123518}, \href {https://ui.adsabs.harvard.edu/abs/2023PhRvD.108l3518L} {108, 123518}

\bibitem[\protect\citeauthoryear{{Macci{\`o}}, {Dutton}, {van den Bosch}, {Moore}, {Potter}  \& {Stadel}}{{Macci{\`o}} et~al.}{2007}]{Macci_2007}
{Macci{\`o}} A.~V.,  {Dutton} A.~A.,  {van den Bosch} F.~C.,  {Moore} B.,  {Potter} D.,   {Stadel} J.,  2007, \mn@doi [\mnras] {10.1111/j.1365-2966.2007.11720.x}, \href {https://ui.adsabs.harvard.edu/abs/2007MNRAS.378...55M} {378, 55}

\bibitem[\protect\citeauthoryear{Mandelbaum et~al.,}{Mandelbaum et~al.}{2005}]{Mandelbaum2005}
Mandelbaum R.,  et~al., 2005, \mn@doi [Monthly Notices of the Royal Astronomical Society] {10.1111/j.1365-2966.2005.09282.x}, 361, 1287

\bibitem[\protect\citeauthoryear{Mandelbaum, Seljak, Kauffmann, Hirata  \& Brinkmann}{Mandelbaum et~al.}{2006}]{Mandelbaum2006}
Mandelbaum R.,  Seljak U.,  Kauffmann G.,  Hirata C.~M.,   Brinkmann J.,  2006, \mn@doi [Monthly Notices of the Royal Astronomical Society] {10.1111/j.1365-2966.2006.10156.x}, 368, 715

\bibitem[\protect\citeauthoryear{Mandelbaum, Slosar, Baldauf, Seljak, Hirata, Nakajima, Reyes  \& Smith}{Mandelbaum et~al.}{2013}]{mandelbaum2013}
Mandelbaum R.,  Slosar A.,  Baldauf T.,  Seljak U.,  Hirata C.~M.,  Nakajima R.,  Reyes R.,   Smith R.~E.,  2013, \mn@doi [Monthly Notices of the Royal Astronomical Society] {10.1093/mnras/stt572}, 432, 1544

\bibitem[\protect\citeauthoryear{{Mandelbaum}, {Wang}, {Zu}, {White}, {Henriques}  \& {More}}{{Mandelbaum} et~al.}{2016}]{Mandelbaum2016}
{Mandelbaum} R.,  {Wang} W.,  {Zu} Y.,  {White} S.,  {Henriques} B.,   {More} S.,  2016, \mn@doi [\mnras] {10.1093/mnras/stw188}, \href {https://ui.adsabs.harvard.edu/abs/2016MNRAS.457.3200M} {457, 3200}

\bibitem[\protect\citeauthoryear{Mandelbaum et~al.,}{Mandelbaum et~al.}{2018}]{Mandelbaum2018b}
Mandelbaum R.,  et~al., 2018, \mn@doi [Monthly Notices of the Royal Astronomical Society] {10.1093/mnras/sty2420}, 481, 3170

\bibitem[\protect\citeauthoryear{Miller}{Miller}{1974}]{Miller1974}
Miller R.~G.,  1974, \mn@doi [Biometrika] {10.1093/biomet/61.1.1}, 61, 1

\bibitem[\protect\citeauthoryear{{Miyatake}, {More}, {Takada}, {Spergel}, {Mandelbaum}, {Rykoff}  \& {Rozo}}{{Miyatake} et~al.}{2016}]{Miyatake2016}
{Miyatake} H.,  {More} S.,  {Takada} M.,  {Spergel} D.~N.,  {Mandelbaum} R.,  {Rykoff} E.~S.,   {Rozo} E.,  2016, \mn@doi [\prl] {10.1103/PhysRevLett.116.041301}, \href {https://ui.adsabs.harvard.edu/abs/2016PhRvL.116d1301M} {116, 041301}

\bibitem[\protect\citeauthoryear{{Miyatake} et~al.,}{{Miyatake} et~al.}{2023}]{Miyatake2023}
{Miyatake} H.,  et~al., 2023, \mn@doi [\prd] {10.1103/PhysRevD.108.123517}, \href {https://ui.adsabs.harvard.edu/abs/2023PhRvD.108l3517M} {108, 123517}

\bibitem[\protect\citeauthoryear{{Miyazaki} et~al.,}{{Miyazaki} et~al.}{2012}]{miyazaki2012}
{Miyazaki} S.,  et~al., 2012, in {McLean} I.~S.,  {Ramsay} S.~K.,   {Takami} H.,  eds,  Society of Photo-Optical Instrumentation Engineers (SPIE) Conference Series Vol. 8446, Ground-based and Airborne Instrumentation for Astronomy IV. p. 84460Z, \mn@doi{10.1117/12.926844}

\bibitem[\protect\citeauthoryear{Miyazaki et~al.,}{Miyazaki et~al.}{2015}]{Miyazaki2015}
Miyazaki S.,  et~al., 2015, \mn@doi [The Astrophysical Journal] {10.1088/0004-637x/807/1/22}, 807, 22

\bibitem[\protect\citeauthoryear{Miyazaki et~al.,}{Miyazaki et~al.}{2017}]{miyazaki2017}
Miyazaki S.,  et~al., 2017, \mn@doi [Publications of the Astronomical Society of Japan] {10.1093/pasj/psx120}, 70

\bibitem[\protect\citeauthoryear{{Moore}, {Katz}, {Lake}, {Dressler}  \& {Oemler}}{{Moore} et~al.}{1996}]{Moore1996}
{Moore} B.,  {Katz} N.,  {Lake} G.,  {Dressler} A.,   {Oemler} A.,  1996, \mn@doi [\nat] {10.1038/379613a0}, \href {https://ui.adsabs.harvard.edu/abs/1996Natur.379..613M} {379, 613}

\bibitem[\protect\citeauthoryear{{Moore}, {Lake}  \& {Katz}}{{Moore} et~al.}{1998}]{Moore1998}
{Moore} B.,  {Lake} G.,   {Katz} N.,  1998, \mn@doi [\apj] {10.1086/305264}, \href {https://ui.adsabs.harvard.edu/abs/1998ApJ...495..139M} {495, 139}

\bibitem[\protect\citeauthoryear{{Moraes} et~al.,}{{Moraes} et~al.}{2014}]{Moraes2014}
{Moraes} B.,  et~al., 2014, in Revista Mexicana de Astronomia y Astrofisica Conference Series. pp 202--203

\bibitem[\protect\citeauthoryear{More, van~den Bosch, Cacciato, Skibba, Mo  \& Yang}{More et~al.}{2010}]{More2011}
More S.,  van~den Bosch F.~C.,  Cacciato M.,  Skibba R.,  Mo H.~J.,   Yang X.,  2010, \mn@doi [Monthly Notices of the Royal Astronomical Society] {10.1111/j.1365-2966.2010.17436.x}, 410, 210

\bibitem[\protect\citeauthoryear{{More} et~al.,}{{More} et~al.}{2016}]{More_2016}
{More} S.,  et~al., 2016, \mn@doi [\apj] {10.3847/0004-637X/825/1/39}, \href {https://ui.adsabs.harvard.edu/abs/2016ApJ...825...39M} {825, 39}

\bibitem[\protect\citeauthoryear{{More} et~al.,}{{More} et~al.}{2023}]{More2023}
{More} S.,  et~al., 2023, \mn@doi [\prd] {10.1103/PhysRevD.108.123520}, \href {https://ui.adsabs.harvard.edu/abs/2023PhRvD.108l3520M} {108, 123520}

\bibitem[\protect\citeauthoryear{{Moster}, {Naab}  \& {White}}{{Moster} et~al.}{2013}]{Moster2013}
{Moster} B.~P.,  {Naab} T.,   {White} S. D.~M.,  2013, \mn@doi [\mnras] {10.1093/mnras/sts261}, \href {https://ui.adsabs.harvard.edu/abs/2013MNRAS.428.3121M} {428, 3121}

\bibitem[\protect\citeauthoryear{{Murata}, {Nishimichi}, {Takada}, {Miyatake}, {Shirasaki}, {More}, {Takahashi}  \& {Osato}}{{Murata} et~al.}{2018}]{Murata2018}
{Murata} R.,  {Nishimichi} T.,  {Takada} M.,  {Miyatake} H.,  {Shirasaki} M.,  {More} S.,  {Takahashi} R.,   {Osato} K.,  2018, \mn@doi [\apj] {10.3847/1538-4357/aaaab8}, \href {https://ui.adsabs.harvard.edu/abs/2018ApJ...854..120M} {854, 120}

\bibitem[\protect\citeauthoryear{{Nadler} et~al.,}{{Nadler} et~al.}{2023}]{Nadler2023}
{Nadler} E.~O.,  et~al., 2023, \mn@doi [\apj] {10.3847/1538-4357/acb68c}, \href {https://ui.adsabs.harvard.edu/abs/2023ApJ...945..159N} {945, 159}

\bibitem[\protect\citeauthoryear{{Nagai} \& {Kravtsov}}{{Nagai} \& {Kravtsov}}{2005}]{Nagai2005}
{Nagai} D.,  {Kravtsov} A.~V.,  2005, \mn@doi [\apj] {10.1086/426016}, \href {https://ui.adsabs.harvard.edu/abs/2005ApJ...618..557N} {618, 557}

\bibitem[\protect\citeauthoryear{{Navarro}, {Frenk}  \& {White}}{{Navarro} et~al.}{1996}]{Navarro1997}
{Navarro} J.~F.,  {Frenk} C.~S.,   {White} S. D.~M.,  1996, \mn@doi [\apj] {10.1086/177173}, \href {https://ui.adsabs.harvard.edu/abs/1996ApJ...462..563N} {462, 563}

\bibitem[\protect\citeauthoryear{Niemiec et~al.,}{Niemiec et~al.}{2017}]{Niemiec2017}
Niemiec A.,  et~al., 2017, \mn@doi [Monthly Notices of the Royal Astronomical Society] {10.1093/mnras/stx1667}, 471, 1153

\bibitem[\protect\citeauthoryear{{Niemiec}, {Jullo}, {Giocoli}, {Limousin}  \& {Jauzac}}{{Niemiec} et~al.}{2019}]{Niemiec2019}
{Niemiec} A.,  {Jullo} E.,  {Giocoli} C.,  {Limousin} M.,   {Jauzac} M.,  2019, \mn@doi [\mnras] {10.1093/mnras/stz1318}, \href {https://ui.adsabs.harvard.edu/abs/2019MNRAS.487..653N} {487, 653}

\bibitem[\protect\citeauthoryear{{Nightingale} et~al.,}{{Nightingale} et~al.}{2022}]{Nightingale2022}
{Nightingale} J.~W.,  et~al., 2022, \mn@doi [arXiv e-prints] {10.48550/arXiv.2209.10566}, \href {https://ui.adsabs.harvard.edu/abs/2022arXiv220910566N} {p. arXiv:2209.10566}

\bibitem[\protect\citeauthoryear{{Nusser}}{{Nusser}}{2024}]{Nusser2024}
{Nusser} A.,  2024, \mn@doi [arXiv e-prints] {10.48550/arXiv.2402.18942}, \href {https://ui.adsabs.harvard.edu/abs/2024arXiv240218942N} {p. arXiv:2402.18942}

\bibitem[\protect\citeauthoryear{{Pastor Mira}, {Hilbert}, {Hartlap}  \& {Schneider}}{{Pastor Mira} et~al.}{2011}]{Poster_mira2011}
{Pastor Mira} E.,  {Hilbert} S.,  {Hartlap} J.,   {Schneider} P.,  2011, \mn@doi [\aap] {10.1051/0004-6361/201116851}, \href {https://ui.adsabs.harvard.edu/abs/2011A&A...531A.169P} {531, A169}

\bibitem[\protect\citeauthoryear{{Planck Collaboration} et~al.,}{{Planck Collaboration} et~al.}{2014}]{Plank2014}
{Planck Collaboration} et~al., 2014, \mn@doi [\aap] {10.1051/0004-6361/201321591}, \href {https://ui.adsabs.harvard.edu/abs/2014A&A...571A..16P} {571, A16}

\bibitem[\protect\citeauthoryear{{Rhee}, {Smith}, {Choi}, {Yi}, {Jaff{\'e}}, {Candlish}  \& {S{\'a}nchez-J{\'a}nssen}}{{Rhee} et~al.}{2017}]{Rhee2017}
{Rhee} J.,  {Smith} R.,  {Choi} H.,  {Yi} S.~K.,  {Jaff{\'e}} Y.,  {Candlish} G.,   {S{\'a}nchez-J{\'a}nssen} R.,  2017, \mn@doi [\apj] {10.3847/1538-4357/aa6d6c}, \href {https://ui.adsabs.harvard.edu/abs/2017ApJ...843..128R} {843, 128}

\bibitem[\protect\citeauthoryear{{Rodriguez-Puebla}}{{Rodriguez-Puebla}}{2024}]{Puebla2024}
{Rodriguez-Puebla} A.,  2024, \mn@doi [arXiv e-prints] {10.48550/arXiv.2404.10801}, \href {https://ui.adsabs.harvard.edu/abs/2024arXiv240410801R} {p. arXiv:2404.10801}

\bibitem[\protect\citeauthoryear{{Rodr{\'\i}guez-Puebla}, {Drory}  \& {Avila-Reese}}{{Rodr{\'\i}guez-Puebla} et~al.}{2012}]{Rodr_puebla2012}
{Rodr{\'\i}guez-Puebla} A.,  {Drory} N.,   {Avila-Reese} V.,  2012, \mn@doi [\apj] {10.1088/0004-637X/756/1/2}, \href {https://ui.adsabs.harvard.edu/abs/2012ApJ...756....2R} {756, 2}

\bibitem[\protect\citeauthoryear{{Rodr{\'\i}guez-Puebla}, {Avila-Reese}  \& {Drory}}{{Rodr{\'\i}guez-Puebla} et~al.}{2013}]{Rodr_puebla2013}
{Rodr{\'\i}guez-Puebla} A.,  {Avila-Reese} V.,   {Drory} N.,  2013, \mn@doi [\apj] {10.1088/0004-637X/767/1/92}, \href {https://ui.adsabs.harvard.edu/abs/2013ApJ...767...92R} {767, 92}

\bibitem[\protect\citeauthoryear{{Rodr{\'\i}guez-Puebla}, {Behroozi}, {Primack}, {Klypin}, {Lee}  \& {Hellinger}}{{Rodr{\'\i}guez-Puebla} et~al.}{2016}]{Rodr_puebla2016}
{Rodr{\'\i}guez-Puebla} A.,  {Behroozi} P.,  {Primack} J.,  {Klypin} A.,  {Lee} C.,   {Hellinger} D.,  2016, \mn@doi [\mnras] {10.1093/mnras/stw1705}, \href {https://ui.adsabs.harvard.edu/abs/2016MNRAS.462..893R} {462, 893}

\bibitem[\protect\citeauthoryear{{Rozo} \& {Rykoff}}{{Rozo} \& {Rykoff}}{2014}]{Rozo_2014}
{Rozo} E.,  {Rykoff} E.~S.,  2014, \mn@doi [\apj] {10.1088/0004-637X/783/2/80}, \href {https://ui.adsabs.harvard.edu/abs/2014ApJ...783...80R} {783, 80}

\bibitem[\protect\citeauthoryear{{Rykoff} et~al.,}{{Rykoff} et~al.}{2014}]{Rykoff2014}
{Rykoff} E.~S.,  et~al., 2014, \mn@doi [\apj] {10.1088/0004-637X/785/2/104}, \href {https://ui.adsabs.harvard.edu/abs/2014ApJ...785..104R} {785, 104}

\bibitem[\protect\citeauthoryear{{Rykoff} et~al.,}{{Rykoff} et~al.}{2016}]{Rykoff2016}
{Rykoff} E.~S.,  et~al., 2016, \mn@doi [\apjs] {10.3847/0067-0049/224/1/1}, \href {https://ui.adsabs.harvard.edu/abs/2016ApJS..224....1R} {224, 1}

\bibitem[\protect\citeauthoryear{{Sand} et~al.,}{{Sand} et~al.}{2012}]{Sand2012}
{Sand} D.~J.,  et~al., 2012, \mn@doi [\apj] {10.1088/0004-637X/746/2/163}, \href {https://ui.adsabs.harvard.edu/abs/2012ApJ...746..163S} {746, 163}

\bibitem[\protect\citeauthoryear{{Schramm} \& {Kayser}}{{Schramm} \& {Kayser}}{1995}]{Schramm_Kayser_1995}
{Schramm} T.,  {Kayser} R.,  1995, \aap, \href {https://ui.adsabs.harvard.edu/abs/1995A&A...299....1S} {299, 1}

\bibitem[\protect\citeauthoryear{{Seitz} \& {Schneider}}{{Seitz} \& {Schneider}}{1997}]{Seitz_Schneider_1997}
{Seitz} C.,  {Schneider} P.,  1997, \aap, \href {https://ui.adsabs.harvard.edu/abs/1997A&A...318..687S} {318, 687}

\bibitem[\protect\citeauthoryear{{Sif{\'o}n} \& {Han}}{{Sif{\'o}n} \& {Han}}{2024}]{Sifon2024}
{Sif{\'o}n} C.,  {Han} J.,  2024, \mn@doi [\aap] {10.1051/0004-6361/202348980}, \href {https://ui.adsabs.harvard.edu/abs/2024A&A...686A.163S} {686, A163}

\bibitem[\protect\citeauthoryear{{Sif{\'o}n}, {Herbonnet}, {Hoekstra}, {van der Burg}  \& {Viola}}{{Sif{\'o}n} et~al.}{2018}]{Sifon2018}
{Sif{\'o}n} C.,  {Herbonnet} R.,  {Hoekstra} H.,  {van der Burg} R. F.~J.,   {Viola} M.,  2018, \mn@doi [\mnras] {10.1093/mnras/sty1161}, \href {https://ui.adsabs.harvard.edu/abs/2018MNRAS.478.1244S} {478, 1244}

\bibitem[\protect\citeauthoryear{Sifón et~al.,}{Sifón et~al.}{2015}]{Sifon2015}
Sifón C.,  et~al., 2015, \mn@doi [Monthly Notices of the Royal Astronomical Society] {10.1093/mnras/stv2051}, 454, 3938

\bibitem[\protect\citeauthoryear{{Simet}, {McClintock}, {Mandelbaum}, {Rozo}, {Rykoff}, {Sheldon}  \& {Wechsler}}{{Simet} et~al.}{2017}]{Simet2017}
{Simet} M.,  {McClintock} T.,  {Mandelbaum} R.,  {Rozo} E.,  {Rykoff} E.,  {Sheldon} E.,   {Wechsler} R.~H.,  2017, \mn@doi [\mnras] {10.1093/mnras/stw3250}, \href {https://ui.adsabs.harvard.edu/abs/2017MNRAS.466.3103S} {466, 3103}

\bibitem[\protect\citeauthoryear{{Springel}, {White}, {Tormen}  \& {Kauffmann}}{{Springel} et~al.}{2001}]{Springel2001}
{Springel} V.,  {White} S. D.~M.,  {Tormen} G.,   {Kauffmann} G.,  2001, \mn@doi [\mnras] {10.1046/j.1365-8711.2001.04912.x}, \href {https://ui.adsabs.harvard.edu/abs/2001MNRAS.328..726S} {328, 726}

\bibitem[\protect\citeauthoryear{{Sugiyama} et~al.,}{{Sugiyama} et~al.}{2023}]{Sugiyama_2023}
{Sugiyama} S.,  et~al., 2023, \mn@doi [\prd] {10.1103/PhysRevD.108.123521}, \href {https://ui.adsabs.harvard.edu/abs/2023PhRvD.108l3521S} {108, 123521}

\bibitem[\protect\citeauthoryear{{Sunayama} \& {More}}{{Sunayama} \& {More}}{2019}]{Sunayama2019}
{Sunayama} T.,  {More} S.,  2019, \mn@doi [\mnras] {10.1093/mnras/stz2832}, \href {https://ui.adsabs.harvard.edu/abs/2019MNRAS.490.4945S} {490, 4945}

\bibitem[\protect\citeauthoryear{{Tanaka}}{{Tanaka}}{2015}]{Tanaka2015}
{Tanaka} M.,  2015, \mn@doi [\apj] {10.1088/0004-637X/801/1/20}, \href {https://ui.adsabs.harvard.edu/abs/2015ApJ...801...20T} {801, 20}

\bibitem[\protect\citeauthoryear{{Tanaka} et~al.,}{{Tanaka} et~al.}{2018}]{Tanaka2018}
{Tanaka} M.,  et~al., 2018, \mn@doi [\pasj] {10.1093/pasj/psx077}, \href {https://ui.adsabs.harvard.edu/abs/2018PASJ...70S...9T} {70, S9}

\bibitem[\protect\citeauthoryear{Tormen, Diaferio  \& Syer}{Tormen et~al.}{1998}]{Tormen_1998}
Tormen G.,  Diaferio A.,   Syer D.,  1998, \mn@doi [Monthly Notices of the Royal Astronomical Society] {10.1046/j.1365-8711.1998.01775.x}, 299, 728

\bibitem[\protect\citeauthoryear{Tormen, Moscardini  \& Yoshida}{Tormen et~al.}{2004}]{Tormen2004}
Tormen G.,  Moscardini L.,   Yoshida N.,  2004, \mn@doi [Monthly Notices of the Royal Astronomical Society] {10.1111/j.1365-2966.2004.07736.x}, 350, 1397

\bibitem[\protect\citeauthoryear{{\VAN{VanDenBosch}{Van den}{van den}}~Bosch, Tormen  \& Giocoli}{{\VAN{VanDenBosch}{Van den}{van den}}~Bosch et~al.}{2005}]{Bosch2005}
{\VAN{VanDenBosch}{Van den}{van den}}~Bosch F.~C.,  Tormen G.,   Giocoli C.,  2005, \mn@doi [Monthly Notices of the Royal Astronomical Society] {10.1111/j.1365-2966.2005.08964.x}, 359, 1029

\bibitem[\protect\citeauthoryear{{\VAN{VanUitert}{Van}{van}}~Uitert, {Hoekstra}, {Velander}, {Gilbank}, {Gladders}  \& {Yee}}{{\VAN{VanUitert}{Van}{van}}~Uitert et~al.}{2011}]{van_uitert2011}
{\VAN{VanUitert}{Van}{van}}~Uitert E.,  {Hoekstra} H.,  {Velander} M.,  {Gilbank} D.~G.,  {Gladders} M.~D.,   {Yee} H.~K.~C.,  2011, \mn@doi [\aap] {10.1051/0004-6361/201117308}, \href {https://ui.adsabs.harvard.edu/abs/2011A&A...534A..14V} {534, A14}

\bibitem[\protect\citeauthoryear{{\VAN{VanUitert}{Van}{van}}~Uitert et~al.,}{{\VAN{VanUitert}{Van}{van}}~Uitert et~al.}{2016}]{van_uitert2016}
{\VAN{VanUitert}{Van}{van}}~Uitert E.,  et~al., 2016, \mn@doi [Monthly Notices of the Royal Astronomical Society] {10.1093/mnras/stw747}, 459, 3251

\bibitem[\protect\citeauthoryear{{Vegetti} \& {Koopmans}}{{Vegetti} \& {Koopmans}}{2009}]{Vegetti2009}
{Vegetti} S.,  {Koopmans} L.~V.~E.,  2009, \mn@doi [\mnras] {10.1111/j.1365-2966.2008.14005.x}, \href {https://ui.adsabs.harvard.edu/abs/2009MNRAS.392..945V} {392, 945}

\bibitem[\protect\citeauthoryear{{Vegetti}, {Koopmans}, {Bolton}, {Treu}  \& {Gavazzi}}{{Vegetti} et~al.}{2010}]{Vegetti2010}
{Vegetti} S.,  {Koopmans} L.~V.~E.,  {Bolton} A.,  {Treu} T.,   {Gavazzi} R.,  2010, \mn@doi [\mnras] {10.1111/j.1365-2966.2010.16865.x}, \href {https://ui.adsabs.harvard.edu/abs/2010MNRAS.408.1969V} {408, 1969}

\bibitem[\protect\citeauthoryear{{Vegetti}, {Lagattuta}, {McKean}, {Auger}, {Fassnacht}  \& {Koopmans}}{{Vegetti} et~al.}{2012}]{Vegetti2012}
{Vegetti} S.,  {Lagattuta} D.~J.,  {McKean} J.~P.,  {Auger} M.~W.,  {Fassnacht} C.~D.,   {Koopmans} L.~V.~E.,  2012, \mn@doi [\nat] {10.1038/nature10669}, \href {https://ui.adsabs.harvard.edu/abs/2012Natur.481..341V} {481, 341}

\bibitem[\protect\citeauthoryear{Velander et~al.,}{Velander et~al.}{2013}]{valander2014}
Velander M.,  et~al., 2013, \mn@doi [Monthly Notices of the Royal Astronomical Society] {10.1093/mnras/stt2013}, 437, 2111

\bibitem[\protect\citeauthoryear{Vogelsberger et~al.,}{Vogelsberger et~al.}{2014}]{Vogelsberger2014}
Vogelsberger M.,  et~al., 2014, \mn@doi [Monthly Notices of the Royal Astronomical Society] {10.1093/mnras/stu1536}, 444, 1518

\bibitem[\protect\citeauthoryear{{Wang} et~al.,}{{Wang} et~al.}{2024}]{Wang2024}
{Wang} C.,  et~al., 2024, \mn@doi [\mnras] {10.1093/mnras/stae121}, \href {https://ui.adsabs.harvard.edu/abs/2024MNRAS.528.2728W} {528, 2728}

\bibitem[\protect\citeauthoryear{{Wright} \& {Brainerd}}{{Wright} \& {Brainerd}}{2000}]{Wright2000}
{Wright} C.~O.,  {Brainerd} T.~G.,  2000, \mn@doi [\apj] {10.1086/308744}, \href {https://ui.adsabs.harvard.edu/abs/2000ApJ...534...34W} {534, 34}

\bibitem[\protect\citeauthoryear{Xie \& Gao}{Xie \& Gao}{2015}]{Xie2015}
Xie L.,  Gao L.,  2015, \mn@doi [Monthly Notices of the Royal Astronomical Society] {10.1093/mnras/stv2077}, 454, 1697

\bibitem[\protect\citeauthoryear{Yang, Mo, van~den Bosch, Jing, Weinmann  \& Meneghetti}{Yang et~al.}{2006}]{Yang2006}
Yang X.,  Mo H.~J.,  van~den Bosch F.~C.,  Jing Y.~P.,  Weinmann S.~M.,   Meneghetti M.,  2006, \mn@doi [Monthly Notices of the Royal Astronomical Society] {10.1111/j.1365-2966.2006.11091.x}, 373, 1159

\bibitem[\protect\citeauthoryear{{Yuan}, {Hadzhiyska}, {Bose}  \& {Eisenstein}}{{Yuan} et~al.}{2022}]{Yuan2022}
{Yuan} S.,  {Hadzhiyska} B.,  {Bose} S.,   {Eisenstein} D.~J.,  2022, \mn@doi [\mnras] {10.1093/mnras/stac830}, \href {https://ui.adsabs.harvard.edu/abs/2022MNRAS.512.5793Y} {512, 5793}

\bibitem[\protect\citeauthoryear{Zhao}{Zhao}{2004}]{Zhao2004}
Zhao H.,  2004, \mn@doi [Monthly Notices of the Royal Astronomical Society] {10.1111/j.1365-2966.2004.07835.x}, 351, 891

\bibitem[\protect\citeauthoryear{{Zhu}, {Avestruz}  \& {Gnedin}}{{Zhu} et~al.}{2020}]{Zhu2020}
{Zhu} H.,  {Avestruz} C.,   {Gnedin} N.~Y.,  2020, \mn@doi [\apj] {10.3847/1538-4357/aba26d}, \href {https://ui.adsabs.harvard.edu/abs/2020ApJ...899..137Z} {899, 137}

\bibitem[\protect\citeauthoryear{{Zu} \& {Mandelbaum}}{{Zu} \& {Mandelbaum}}{2015}]{zu2015}
{Zu} Y.,  {Mandelbaum} R.,  2015, \mn@doi [\mnras] {10.1093/mnras/stv2062}, \href {https://ui.adsabs.harvard.edu/abs/2015MNRAS.454.1161Z} {454, 1161}

\makeatother
\end{thebibliography}
\section*{Appendix}
\appendix
\renewcommand{\thefigure}{A\arabic{figure}}
\setcounter{figure}{0}
\renewcommand{\thetable}{A\arabic{table}}
\setcounter{table}{0}

\begin{figure*}
        \begin{subfigure}[b]{0.5\textwidth}
        \includegraphics[width=0.99\columnwidth]{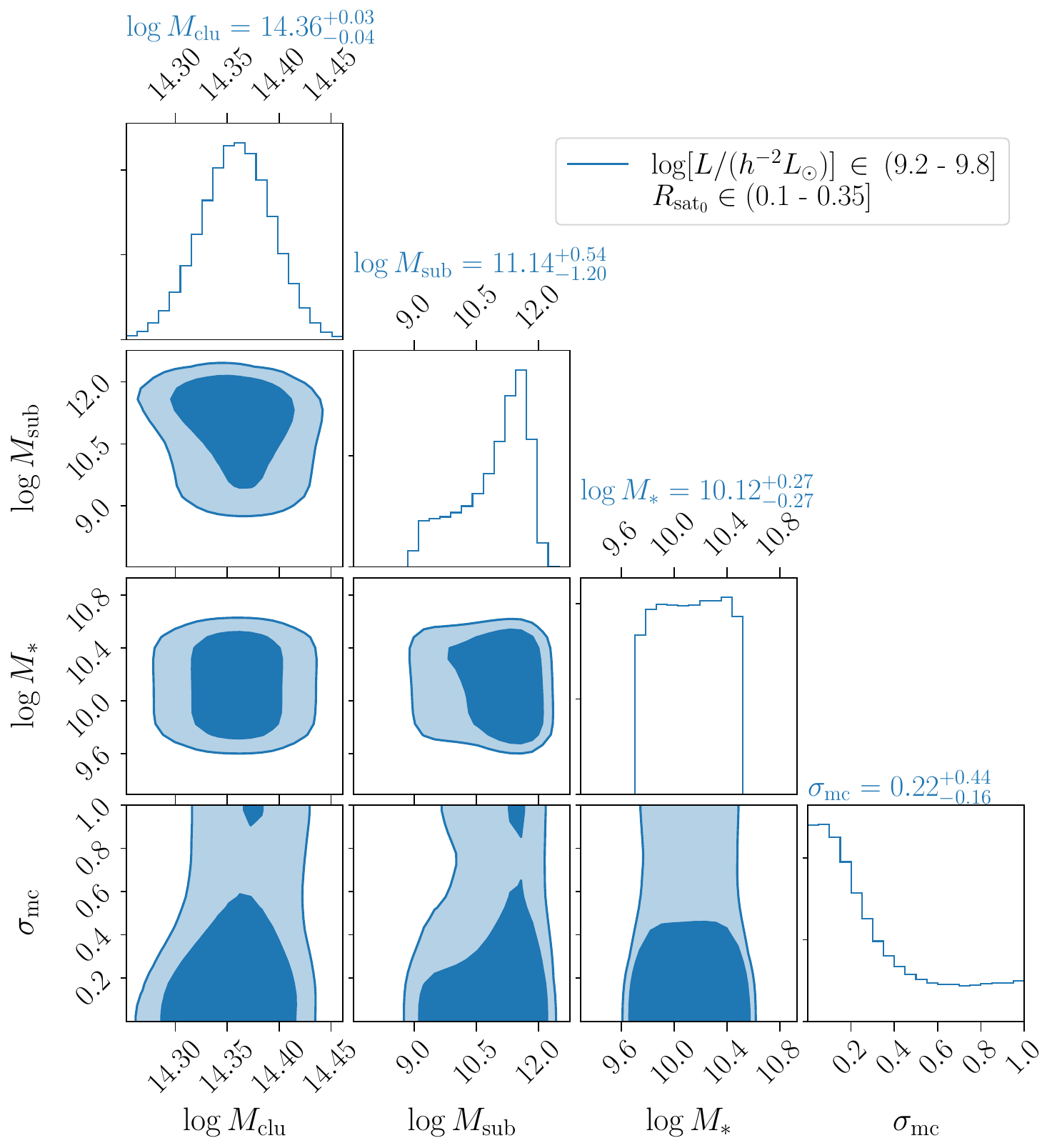}         \end{subfigure}%
        \begin{subfigure}[b]{0.5\textwidth}
            \includegraphics[width=0.99\columnwidth]{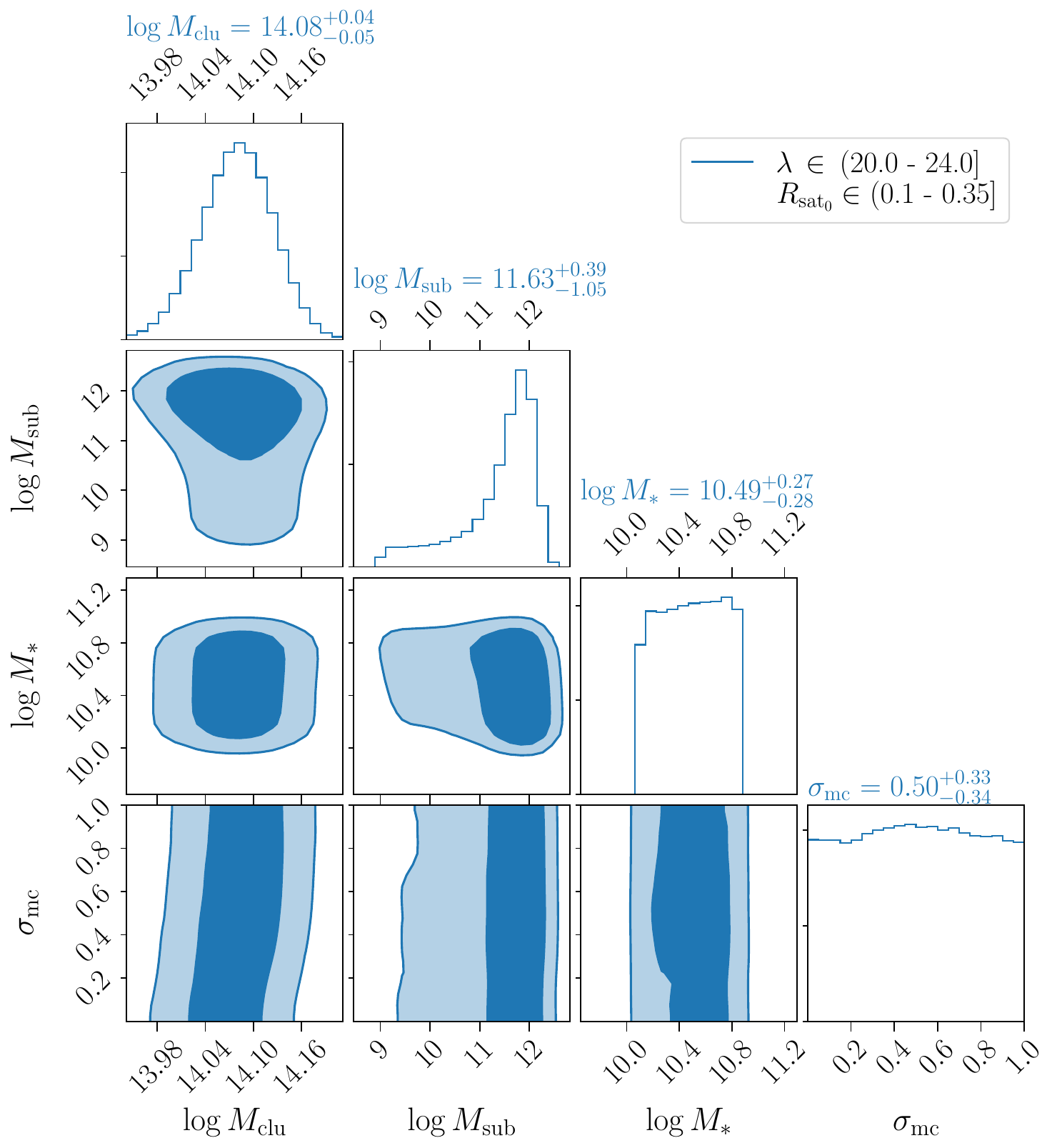}
        \end{subfigure}
        \caption{ The figure shows the posterior distribution of the model parameters used in our MCMC sampling. The left figure represents the 2-D histograms of the model parameters for the signal fits in one of the luminosity-based bins, while the right one corresponds to an analysis for one of the richness-based bins. The figures are representative of what we observe in other bins, too. In Table \ref{Pub3_table:infered_parameters}, we summarise the posterior statistics for all the bins used in our analysis.}
        \label{Pub3_fig:Posterior_distribution}
\end{figure*}
\bsp	
\label{Pub3_lastpage}
\end{document}